\documentclass[pdflatex,sn-mathphys-ay]{sn-jnl}

\usepackage{mathrsfs}
\usepackage{bibunits}
\usepackage{graphicx}
\usepackage{multirow}
\usepackage{amsmath,amssymb,amsfonts}
\usepackage{amsthm}
\usepackage[title]{appendix}
\usepackage{textcomp}
\usepackage{makecell} 
\usepackage{booktabs}
\usepackage{siunitx}
\usepackage{comment}
\usepackage{colortbl}
\usepackage{tabularx}
\usepackage{pifont}
\usepackage{threeparttable}
\usepackage{array}
\usepackage[normalem]{ulem}
\usepackage{subcaption}
\usepackage{lineno}
\usepackage[table]{xcolor}
\usepackage{natbib}
\usepackage{etoolbox}

\makeatletter

\patchcmd{\@maketitle}
    {\newpage\null}
    {\newpage
     \thispagestyle{empty}
     \enlargethispage{25mm}
     \null}
    {}
    {\PackageWarning{layout}{Patch del frontespizio non riuscita}}

\makeatother

\defaultbibliographystyle{sn-mathphys-ay}

\DeclareCaptionFont{customcaption}{\fontsize{8.6}{9.4}\selectfont}
 
\theoremstyle{thmstyleone}

\theoremstyle{thmstyletwo}

\theoremstyle{thmstylethree}

\newcommand{\misscit}[1]{\textcolor{red}{[cit?]}}

\hypersetup{hidelinks}

\definecolor{lightgray}{gray}{0.95}
\definecolor{lightergray}{gray}{0.97}
\definecolor{good}{RGB}{46,139,87}
\definecolor{bad}{RGB}{178,34,34}
\definecolor{neutral}{RGB}{70,70,70}
\definecolor{lightgood}{RGB}{220,245,235}
\definecolor{lightbad}{RGB}{250,225,225}
\definecolor{lightblue}{RGB}{225,238,250}
\definecolor{takeawayBlue}{RGB}{239,247,252}
\definecolor{takeawayGreen}{RGB}{240,248,240}
\definecolor{takeawayYellow}{RGB}{252,248,232}

\begin{document}
\begin{bibunit}[sn-mathphys-ay]

\title[Oxygenation and spatial heterogeneity shape radiotherapy protocol ranking through phenotypic adaptation]
{Oxygenation and spatial heterogeneity shape radiotherapy protocol ranking through phenotypic adaptation}

% \subtitle{%
%     \normalfont\fontsize{9}{11}\selectfont
%     DOI: \href{https://doi.org/10.1007/s11538-026-01758-5}
%     {10.1007/s11538-026-01758-5}
% }

% ORCID Francesco Albanese: 0009-0009-2588-4168
\author*[1]{\fnm{Francesco} \sur{Albanese}}\email{francesco\_albanese@polito.it}

% ORCID Giulia Chiari: 0000-0002-4432-773X
\author[2,3]{\fnm{Giulia} \sur{Chiari}}

% ORCID Marcello Edoardo Delitala: 0000-0002-5347-5654
\author[1]{\fnm{Marcello Edoardo} \sur{Delitala}}

\affil*[1]{\orgdiv{Department of Mathematical Sciences ``G. L. Lagrange''}, \orgname{Politecnico di Torino}, \orgaddress{\city{Torino}, \country{Italy}}}

\affil[2]{\orgname{Basque Center for Applied Mathematics (BCAM)}, \orgaddress{\city{Bilbao}, \country{Spain}}}

\affil[3]{\orgdiv{Mathematical Institute}, \orgname{University of Oxford}, \orgaddress{\city{Oxford}, \country{United Kingdom}}}

\abstract{
    Tumor response to radiotherapy is strongly influenced by oxygen availability and phenotypic heterogeneity, yet their combined impact on the relative performance of fractionation schedules remains unclear. Here, we develop a mathematical model that integrates spatial oxygen dynamics with continuous phenotypic adaptation to hypoxia and radiation, and use it to systematically compare radiotherapy protocols under a common normal-tissue toxicity constraint.
    Under spatially uniform oxygenation, we find that alternative fractionation schedules provide little improvement over standard-of-care protocols in normoxic conditions. Under moderate hypoxia, however, a distinct class of protracted schedules with longer inter-fraction intervals substantially increases time-to-progression, in some cases by up to twofold. This regime-dependent benefit is consistent with a shift in the balance between reoxygenation and selection for resistant phenotypes.
    When oxygen delivery is spatially heterogeneous, treatment outcomes depend strongly on the geometric organization of oxygen sources. Even with identical total oxygen supply, different spatial configurations lead to large variability in time-to-progression and can alter the relative ranking of radiotherapy protocols.
    These results show that radiotherapy effectiveness is not an intrinsic property of a treatment schedule alone, but emerges from its interaction with tumor microenvironmental structure and evolutionary dynamics. Incorporating both spatial heterogeneity and phenotypic adaptation may therefore be important for the consistent evaluation and design of fractionation strategies in heterogeneous tumors.
}

\keywords{
Radiotherapy fractionation; Phenotypic and spatial heterogeneity; Radioresistance;
Tumor hypoxia; Evolutionary dynamics; Mathematical oncology}

\maketitle

\section{Introduction}
    Solid tumors are widely recognized as heterogeneous and evolving systems, in which malignant cells interact dynamically with their microenvironment \citep{Hanahan2000}. 
    Among microenvironmental factors, oxygen availability plays a central role in shaping both tumor progression and response to therapy.
    Radiotherapy represents a standard treatment modality \citep{Siegel2023}, whose efficacy is strongly modulated by local oxygen levels. Oxygen enhances radiation-induced DNA damage, an effect quantified through the Oxygen Enhancement Ratio (OER), defined as the dose increase required under hypoxia to match the biological effect observed in normoxia.
    However, oxygen influences treatment response in more than one way. Beyond its direct physico-chemical role in damage fixation, hypoxia induces transcriptional and metabolic reprogramming, promoting aggressive phenotypes that contribute to radioresistance \citep{Beckers2024}. Importantly, experimental studies indicate that OER is not uniform across intratumoral subpopulations. More aggressive cells are characterized by lower intrinsic radiosensitivity \citep{Lagadec2012}.
    Moreover, oxygen distribution within the tumor microenvironment (TME) is typically spatially heterogeneous, owing to the irregular arrangement and perfusion of the tumor vasculature. As a consequence, not only the overall level of oxygenation, but also the geometry of oxygen sources and the resulting spatial gradients may contribute to shaping local radiotherapy response \citep{Hormuth2021}.\\
    \indent Mathematical models of tumor response to radiotherapy combine population growth dynamics with radiobiological descriptions of radiation-induced cell killing. Tumor kinetics are typically represented through proliferation laws incorporating carrying-capacity effects \citep{Zheng2025}, whereas radiation response is described by the linear--quadratic (LQ) formalism, in which the surviving fraction after a dose $d$ is $S(d)=\exp(-\alpha d - \beta d^2)$, with $\alpha$ and $\beta$ denoting radiosensitivity parameters inferred from experimental and clinical studies \citep{vanLeeuwen2018}. The LQ formulation also underlies the definition of biologically effective dose (BED), enabling comparisons across fractionation schedules.
    Experimental and clinical evidence on hypoxia led to extensions of the classical LQ model incorporating oxygen dependence. In particular, radiosensitivity parameters have been rescaled through an OER factor to account for variations in radiation response between normoxic and hypoxic conditions \citep{Wenzl2011}. Alternative approaches have relaxed the assumption of homogeneous radiosensitivity by allowing the parameters $\alpha$ and $\beta$ to vary across tumor cell populations or between patients, thereby representing intrinsic biological heterogeneity \citep{Alfonso2019}.
    Trait-structured models have recently been employed to describe tumor response to radiotherapy in the presence of both phenotypic heterogeneity and microenvironmental modulation \citep{Celora2023, Chiari2023b}. In these formulations, spatial dynamics and continuous phenotypic structure are coupled within a unified PDE system, so that tumor response under irradiation can be interpreted as an eco-evolutionary process in which selection emerges from the redistribution of cells across phenotypic states.\\
    \indent Several mathematical studies have explored alternative radiotherapy fractionation strategies \citep{Prokopiou2015,HenaresMolina2017,Bruningk2021}. To our knowledge, however, the combined impact of spatially heterogeneous oxygenation and continuous phenotypic adaptation on the comparative performance of treatment schedules has not been systematically investigated. In particular, several protocols currently adopted in clinical practice are considered optimal for tumors exhibiting specific cellular and microenvironmental features; however, a mechanistic modeling framework capable of explaining these observed treatment preferences as a consequence of such features is still lacking.\\
    \indent In this context, this study is motivated by the need to understand how the interplay between oxygen spatial structure, evolutionary selection of resistant phenotypes, and normal-tissue toxicity constraints may shape the landscape of admissible fractionation protocols. 
    Are there dose--interval combinations, beyond those commonly used in clinical practice, that may lead to improved treatment outcomes? How does spatial heterogeneity in oxygen delivery influence the relative performance of radiotherapy schedules? Are protocols identified under spatially uniform conditions sufficiently robust, or is detailed characterization of spatial oxygen heterogeneity required for reliable treatment selection?\\
    \indent To address these questions, we develop a phenotypically structured partial differential equation (PS-PDE) model \citep{Lorenzi2025}. Within this framework, we systematically explore the dose--interval plane under biologically effective dose constraints, enabling comparison of fractionation protocols across different oxygenation regimes and tissue tolerance conditions. This modeling strategy aligns with current efforts toward biologically informed radiotherapy design, including multi-omics--guided treatment adaptation \citep{Boldrini2024} and spatially targeted dose painting \citep{Bentzen2011}.\\
    \indent The remainder of the paper is organized as follows. Section~\ref{sec: methods} introduces the mathematical model, including tumor spatio--phenotypic dynamics, oxygen dynamics, and the oxygen- and phenotype-dependent formulation of radiotherapy response. Section~\ref{sec:results} investigates how oxygenation affects the performance of fractionation schedules under a common toxicity constraint, first under spatially uniform oxygen supply and then under heterogeneous oxygen delivery. Particular attention is given to the role of phenotypic adaptation and to how the spatial organization of oxygen supply can alter time-to-progression and protocol ranking. Conclusions and perspectives are discussed in Section~\ref{sec:discussion}.           
\section{Methods}
\label{sec: methods}
    We introduce a spatio--phenotypic model coupling tumor-cell dynamics, oxygen diffusion--consum\-ption, and oxygen- and phenotype-dependent radiotherapy response. Model parameters are listed in Table 1. Simulation details and indicator definitions are reported in the Supporting Information S1 and S2.

    \subsection{Cell Dynamics}
        We consider a portion of biological tissue where the early growth of a tumor mass occurs. The spatio--temporal evolution of the active cancer cell population is described by a density function $n(t,x,u)$, where \(t\) denotes time, \(x \in \Omega_s \subset \mathbb{R}\) represents the spatial position, and \(u \in \Omega_p = [0,1]\) is a continuous phenotypic trait.  
        Although spatial dynamics are modeled along a single coordinate, $n(t,x,u)$ is interpreted as a volumetric cell density. Following the framework introduced in \citet{Chiari2023b}, we model resistance as a continuous variable ranging from fully hypoxia- and radiation-sensitive cells (\(u=0\)) to fully resistant ones (\(u=1\)), with higher values of \(u\) corresponding to increased resistance to both stresses. We emphasize that \(u\) represents an intrinsic phenotypic adaptation state, shaped by selection and non-genetic phenotypic changes, rather than a marker of a cell's current, instantaneous oxygen exposure. A cell with high \(u\) therefore represents a phenotype adapted to hypoxic stress and intrinsically more resistant to radiation, consistently with the well-established association between hypoxia-induced cellular adaptation and radioresistance \citep{Beckers2024}.\\
        \indent The spatio--phenotypic avascular evolution of the tumor cell density is modeled through a Phenotypically Structured Partial Differential Equation (PS-PDE):        
        \begin{equation}
            \partial_t n(t,x,u)
            =
            D_p\,\partial_u^2 n(t,x,u)
            +
            D_s\,\partial_x^2 n(t,x,u)
            +
            \bigl[R(t,x,u)-T(t,x,u)\bigr]\,n(t,x,u)
            \label{eq:population}
        \end{equation}
    
        \noindent where \(D_p\) represents the coefficient of phenotypic diffusion, accounting for stochastic epigenetic changes. The term \(D_s\) corresponds to the linear spatial diffusion coefficient, which describes the random dispersal of cells within the tissue. 
        The function \(R(t,x,u)\) defines the net proliferation rate of the tumor cell population, whereas \(T(t,x,u)\) quantifies the effective rate of cell death induced by radiotherapy. Both terms depend on time, space, and phenotype, reflecting the heterogeneous nature of the tumor response to oxygen availability and radiation exposure.
                
        \subsubsection{Reaction term}
            \label{sec:reaction term}
            The proliferation rate \(R(t,x,u)\) models cell growth as influenced by oxygen availability, phenotypic adaptation, and crowding effects:
            \begin{equation}
            R\bigl(t,x,u\bigr)
            =
            r_0
            +
            \gamma_O\,M\bigl(O(t,x)\bigr)\,(1-u^2)
            -
            \gamma_H\bigl(1-M\bigl(O(t,x)\bigr)\bigr)\,(1-u)^2
            -
            \kappa\,\rho(t,x)
            \label{eq:reaction_term}
            \end{equation}
            
            \noindent where \(r_0\), \(\gamma_O\), \(\gamma_H\), and \(\kappa\) are model parameters, and
            \(O(t,x):[0,t_{end}]\times\Omega_s\to\mathbb{R}_{+}\) denotes the local oxygen concentration.
            
            The first term \(r_0\) represents the baseline proliferation rate of fully resistant cells (\(u=1\)), which sustain a minimal growth rate even under adverse environmental conditions.
            
            The second term provides an additional proliferative contribution, weighted by two effects:
            (i) a Hill-type function enhancing proliferation under normoxic conditions,
            \begin{equation}
                M\bigl(O(t,x)\bigr)
                =
                \frac{O(t,x)^4}{O(t,x)^4+(\alpha_O)^4}
                \label{eq:hill_function}
            \end{equation}
            and (ii) a quadratic dependence on \(u\), expressing that phenotypes less resistant to radiotherapy and less adapted to hypoxia ($u \to 0$) exhibit higher proliferative potential.
            This formulation reflects an energetic trade-off, whereby cells investing metabolic resources in resistance to hypoxia or radiation damage proliferate less efficiently. 
            The Hill exponent is set to \(n_H=4\) as a modeling choice, providing a sharp
            but smooth transition and an effectively saturated oxygen-dependent
            proliferative contribution within the normoxic range. This value is not intended
            to represent a specific molecular cooperativity mechanism. The robustness of
            the results presented below with respect to this choice is assessed in
            Supporting Information S10 by repeating the analysis for different Hill
            coefficients.
            
            The third term penalizes poorly adapted phenotypes under low-oxygen conditions, thereby acting as a selective pressure toward higher-resistance traits. The parameter $\gamma_H$ is set to $\gamma_H = r_0+\gamma_O$, so that under extreme hypoxia the penalty experienced by poorly adapted phenotypes offsets the largest intrinsic growth rate attained under favorable oxygenation.
            
            The quadratic functional form in both the second and third terms is chosen so as to induce a concave fitness landscape: for fixed \(t\) and \(x\), and depending on the oxygen concentration, the reaction term admits a unique global maximum with respect to the phenotypic variable \(u\), as discussed in Supporting Information S3. This modeling assumption is consistent with analogous choices adopted in the literature on PS-PDEs \citep{Lorenzi2025}.
            
            Finally, the crowding term enforces population saturation due to spatial limitations.
            Here, $\rho(t,x)$ denotes the total local cell density,
            \begin{equation}
            \rho(t,x) = \int_{\Omega_p} n(t,x,u)\,du.
            \label{eq:total_density}
            \end{equation}
            For fixed $(t,x,u)$, the net proliferation rate $R(t,x,u)$ is decreasing with respect to the local density $\rho$, thereby limiting local tumor growth as cell density increases. The crowding coefficient $\kappa$ is calibrated in Supporting Information S4 by prescribing \(K\) as an upper-bound carrying capacity under the most favorable proliferative conditions.
            
            Overall, the reaction term defines a dynamic fitness landscape that evolves in time with both the oxygen distribution and the phenotypic composition of the tumor population.        
            
    \subsection{Radiotherapy term}
        \label{subsubsection:RadiotherapyTerm}
        
        Radiation-induced cell death is modeled through a phenotype- and oxygen-dependent extension of the linear--quadratic (LQ) formulation. We write
        \begin{equation}
        \label{eq:T}
        T(t,x,u)
        =
        \bigl[\alpha(O(t,x),u)\, d
        +
        \beta(O(t,x),u)\, d^2
        \bigr]\,
        \delta_{\mathcal T}(t),
        \end{equation}
        where $d$ denotes the dose delivered per fraction (Gy), and $\delta_{\mathcal T}(t)$ is a sum of Dirac delta functions that localize irradiation at prescribed treatment times, as detailed below.
        The coefficients $\alpha(O,u)$ and $\beta(O,u)$ are continuous counterparts of the classical constant LQ parameters and determine the survival fraction $S(d)=\exp \bigl(-T(t,x,u)\big)$ in the standard formulation \citep{McMahon2018}. In our model they depend on both oxygen concentration and phenotype:
        \begin{equation}
        \label{eq:alphabeta_new}
        \begin{aligned}
        \alpha(O,u) &= 
        \dfrac{\tilde{\alpha} + \Delta_\alpha (1-u)^2}{\mathrm{OER}(O)} \\[4pt]
        \beta(O,u)  &= 
        \dfrac{\tilde{\beta} + \Delta_\beta (1-u)^2}{\mathrm{OER}^2(O)}
        \end{aligned}
        \end{equation}
        The quadratic dependence on $(1-u)$ introduces phenotype-specific modulation of radiosensitivity, consistently with patient-specific modeling studies showing that higher proliferation rates are associated with increased radiosensitivity \citep{Rockne2010}. 
        The OER formulation effectively rescales the delivered dose, so that $d/\mathrm{OER}(O)$ represents an oxygen-dependent effective dose, with the quadratic term scaling accordingly as $d^2/\mathrm{OER}^2(O)$.
        The values of $\tilde{\alpha}$, $\tilde{\beta}$, $\Delta_{\alpha}$, and $\Delta_{\beta}$ are reported in Table~\ref{tab:parameters}. Further details on their derivation are provided in Supporting Information S5, where we also explore in depth the rationale underlying Eq.~\eqref{eq:alphabeta_new} and the theoretical implications of this formulation, which are used throughout the results presented in the following sections.
        
        In addition to this intrinsic phenotypic contribution, radiosensitivity is further modulated by the local oxygen concentration: increasing oxygen availability enhances radiation-induced killing through an OER-based rescaling of the LQ parameters \(\alpha(O,u)\) and \(\beta(O,u)\). We adopt the following functional form
        \begin{equation}
            \label{eq:OER}
            \mathrm{OER}(O) = 1 + \frac{(\mathrm{OER}_{\max}-1)\, K_O}{\mathrm{OER}_{\max}\,O + K_O}
        \end{equation}
        in line with the saturation law proposed by Howard-Flanders and Alper~\citeyearpar{HowardFlanders1957}.
        Since the OER rescales the effective delivered dose as \(d/\mathrm{OER}(O)\), the reciprocal factor \(1/\mathrm{OER}(O)\) is the quantity that increases with oxygenation in the radiation-response term. This factor ranges from \(1/\mathrm{OER}_{\max}\) under anoxic conditions \((O\to0)\) to \(1\) under well-oxygenated conditions \((O\to+\infty)\). The parameter \(K_O\) corresponds to the oxygen level at which \(1/\mathrm{OER}(O)\) reaches the midpoint between its anoxic limit \(1/\mathrm{OER}_{\max}\) and its well-oxygenated limit \(1\).
        
        Radiotherapy protocols are characterized by equally spaced isodoses and are
        denoted throughout by the triple $\mathcal P=(d,\Delta,N_f)$, where $d$ is the dose per fraction, $\Delta$ the inter-fraction interval, and $N_f$ the total number of fractions. If treatment starts at time $t^{RT}_0$, the set of irradiation times is $\mathcal T=\{t^{RT}_0+j\Delta \;|\; j=0,\dots,N_f-1\}$.
        The formulation in Eq.~\eqref{eq:T} is equivalent to imposing an instantaneous update of the cell density at each irradiation time $t^{RT}_i$:
        \begin{equation}
        \label{eq:rt_jump_new}
        n(t^{RT,+}_i,x,u)
        =
        S(d)\,
        n(t^{RT,-}_i,x,u),
        \end{equation}

        % ============================================================
        % TABLE 1 
        % ============================================================
        
        \newlength{\ParamW}
        \newlength{\DescW}
        \newlength{\ValueW}
        \newlength{\RefW}
        \newlength{\RowH}
        \newlength{\HeaderH}
        
        \setlength{\HeaderH}{1.60em}
        \setlength{\ParamW}{0.12\textwidth}
        \setlength{\DescW}{0.34\textwidth}
        \setlength{\ValueW}{0.28\textwidth}
        \setlength{\RefW}{0.20\textwidth}
        
        \setlength{\RowH}{2.20em}
        
        \newcommand{\PCell}[1]{%
            \parbox[c][\RowH][c]{\ParamW}{\centering #1}%
        }
        
        \newcommand{\DCell}[1]{%
            \parbox[c][\RowH][c]{\DescW}{\raggedright #1}%
        }
        
        \newcommand{\VCell}[1]{%
            \parbox[c][\RowH][c]{\ValueW}{\centering #1}%
        }
        
        \newcommand{\RCell}[1]{%
            \parbox[c][\RowH][c]{\RefW}{\centering #1}%
        }       
        \clearpage
        \begin{center}
        
            \fontsize{7.0}{7.8}\selectfont
        
            % Poco spazio orizzontale aggiuntivo
            \setlength{\tabcolsep}{1.8pt}
            \renewcommand{\arraystretch}{1}
        
            \begin{tabular}{|
                c|
                c|
                c|
                c|}
        
                \hline        
                \rowcolor{gray!20}
                \parbox[c][\HeaderH][c]{\ParamW}{\centering
                    \fontsize{6.8}{7.1}\selectfont
                    \scalebox{0.93}[1]{\textbf{Parameter}}} &
                \parbox[c][\HeaderH][c]{\DescW}{\centering
                    \fontsize{6.8}{7.1}\selectfont
                    \scalebox{0.93}[1]{\textbf{Description}}} &
                \parbox[c][\HeaderH][c]{\ValueW}{\centering
                    \fontsize{6.8}{7.1}\selectfont
                    \scalebox{0.93}[1]{\textbf{Value (units)}}} &
                \parbox[c][\HeaderH][c]{\RefW}{\centering
                    \fontsize{6.8}{7.1}\selectfont
                    \scalebox{0.93}[1]{\textbf{Ref.}}} \\
                \hline
        
                % ====================================================
                % CELL DYNAMICS
                % ====================================================
        
                \rowcolor{lightblue}
                \PCell{$D_p$} &
                \DCell{Phenotypic diffusion rate} &
                \VCell{$8.64 \cdot 10^{-7}\;(\mathrm{day}^{-1})$} &
                \RCell{M.E.} \\
                \hline
        
                \rowcolor{lightblue}
                \PCell{$D_s$} &
                \DCell{Spatial diffusion rate} &
                \VCell{$3.11 \cdot 10^{-5}\;(\mathrm{cm^2\,day^{-1}})$} &
                \RCell{\citealp{Chiari2023b}} \\
                \hline
        
                \rowcolor{lightblue}
                \PCell{$r_0$} &
                \DCell{Baseline proliferation rate} &
                \VCell{$2 \cdot 10^{-3}\;(\mathrm{day}^{-1})$} &
                \RCell{\citealp{HenaresMolina2017}} \\
                \hline
        
                \rowcolor{lightblue}
                \PCell{$\gamma_O$} &
                \DCell{Oxygen-dependent proliferative gain} &
                \VCell{$1 \cdot 10^{-2}\;(\mathrm{day}^{-1})$} &
                \RCell{\citealp{HenaresMolina2017}} \\
                \hline
        
                \rowcolor{lightblue}
                \PCell{$\gamma_H$} &
                \DCell{Hypoxia-driven selection strength} &
                \VCell{$1.2 \cdot 10^{-2}\;(\mathrm{day}^{-1})$} &
                \RCell{M.E.} \\
                \hline
        
                \rowcolor{lightblue}
                \PCell{$K$} &
                \DCell{Tissue carrying capacity} &
                \VCell{$10^{9}\;(\mathrm{cells\,cm^{-3}})$} &
                \RCell{\citealp{DelMonte2009}} \\
                \hline
        
                \rowcolor{lightblue}
                \PCell{$\kappa$} &
                \DCell{Crowding coefficient} &
                \VCell{{\fontsize{6.4}{6.8}\selectfont
                $1.2 \cdot 10^{-11}\;(\mathrm{cm^3\,cells^{-1}\,day^{-1}})$
                }} &
                \RCell{S4} \\
                \hline
        
                \rowcolor{lightblue}
                \PCell{$\alpha_O$} &
                \DCell{Oxygen half-saturation constant} &
                \VCell{$15.2\;(\mathrm{mmHg})$} &
                \RCell{M.E.} \\
                \hline
        
                % ====================================================
                % RADIOTHERAPY RESPONSE
                % ====================================================
        
                \rowcolor{red!6}
                \PCell{$\tilde{\alpha}$} &
                \DCell{Basal linear radiosensitivity parameter} &
                \VCell{$6 \cdot 10^{-3}\;(\mathrm{Gy}^{-1})$} &
                \RCell{S5} \\
                \hline
        
                \rowcolor{red!6}
                \PCell{$\tilde{\beta}$} &
                \DCell{Basal quadratic radiosensitivity parameter} &
                \VCell{$3.6 \cdot 10^{-3}\;(\mathrm{Gy}^{-2})$} &
                \RCell{S5} \\
                \hline
        
                \rowcolor{red!6}
                \PCell{$\Delta_\alpha$} &
                \DCell{Phenotype-dependent modulation range of $\alpha$} &
                \VCell{$2.34 \cdot 10^{-1}\;(\mathrm{Gy}^{-1})$} &
                \RCell{S5} \\
                \hline
        
                \rowcolor{red!6}
                \PCell{$\Delta_\beta$} &
                \DCell{Phenotype-dependent modulation range of $\beta$} &
                \VCell{$2.52 \cdot 10^{-2}\;(\mathrm{Gy}^{-2})$} &
                \RCell{S5} \\
                \hline
        
                \rowcolor{red!6}
                \PCell{$\mathrm{OER}_{\max}$} &
                \DCell{Maximum oxygen enhancement ratio} &
                \VCell{$3$} &
                \RCell{\citealp{Grimes2015}} \\
                \hline
        
                \rowcolor{red!6}
                \PCell{$K_O$} &
                \DCell{Oxygen level for half-transition of radiosensitivity} &
                \VCell{$3.00\;(\mathrm{mmHg})$} &
                \RCell{{\fontsize{6.2}{6.2}\selectfont\citealp{McKeown2014,Grimes2015}}} \\
                \hline
        
                \rowcolor{red!6}
                \PCell{$\Gamma_{\mathrm{RT}}$} &
                \DCell{Radiotherapy detection threshold} &
                \VCell{$1.3 \cdot 10^{8}\;(\mathrm{cells})$} &
                \RCell{S1} \\
                \hline
        
                % ====================================================
                % OXYGEN KINETICS
                % ====================================================
        
                \rowcolor{green!6}
                \PCell{$D_O$} &
                \DCell{Oxygen diffusion coefficient} &
                \VCell{$8.64 \cdot 10^{-1}\;(\mathrm{cm^2\,day^{-1}})$} &
                \RCell{\citealp{MartinezGonzalez2012}} \\
                \hline
        
                \rowcolor{green!6}
                \PCell{$\lambda_O$} &
                \DCell{Baseline oxygen consumption rate} &
                \VCell{$2.23 \cdot 10^{3}\;(\mathrm{day}^{-1})$} &
                \RCell{S6} \\
                \hline
        
                \rowcolor{green!6}
                \PCell{$\zeta_O$} &
                \DCell{Tumor oxygen excess uptake} &
                \VCell{{\fontsize{6.4}{6.8}\selectfont
                $5.8 \cdot 10^{-5}\;(\mathrm{cm^3\,day^{-1}\,cell^{-1}})$
                }} &
                \RCell{S6} \\
                \hline
        
                \rowcolor{green!6}
                \PCell{$O_M$} &
                \DCell{Normoxia threshold} &
                \VCell{$72.2\;(\mathrm{mmHg})$} &
                \RCell{\citealp{McKeown2014}} \\
                \hline
        
                \rowcolor{green!6}
                \PCell{$O_m$} &
                \DCell{Physiological hypoxia threshold} &
                \VCell{$30.4\;(\mathrm{mmHg})$} &
                \RCell{\citealp{McKeown2014}} \\
                \hline
        
                \rowcolor{green!6}
                \PCell{$O_h$} &
                \DCell{Pathological hypoxia threshold} &
                \VCell{$7.60\;(\mathrm{mmHg})$} &
                \RCell{\citealp{McKeown2014}} \\
                \hline
        
            \end{tabular}
        
            \vspace{0.5mm}
            {
            \captionsetup{font=customcaption}
            \captionof{table}{
            Model parameters. Background colors distinguish parameter groups associated with cell dynamics (blue), radiotherapy response (red), and oxygen kinetics (green). The final column reports the origin of each parameter value, indicating whether it is taken from the literature, derived in a specific section of the paper, or estimated within the present framework as a model estimate (M.E.). Entries labeled with ``S'' refer to the corresponding section of the Supporting Information.
            }
            }
            \label{tab:parameters}
        
        \end{center}
        
        \clearpage

        \noindent where $t^{RT,-}_i$, $t^{RT,+}_i$ denote instants immediately before and after dose delivery.
        Cells lethally damaged by radiation are assumed to be removed instantaneously. Delayed effects (e.g., senescence, necrotic accumulation, and radiation-induced microenvironmental remodeling) as well as sublethal damage repair are not explicitly modeled; accordingly, corrections such as the Lea-Catcheside protraction factor \citep{Brenner2008} are not included.

    \subsection{Oxygen Dynamics}
        The spatio--temporal evolution of the oxygen concentration $O(t,x)$ is described by the following diffusion--consumption equation:
           
        \begin{equation}
            \partial_t O(t,x)
            =
            D_O\,\partial_{x}^2 O(t,x)
            -
            \lambda_O\,O(t,x)
            -
            \zeta_O\,\rho(t,x)\,O(t,x)
            +
            V_O(x)
            \label{eq:oxygen_dynamics}
        \end{equation}
        Here, $D_O$ denotes the oxygen diffusion coefficient in the tissue, and a Fickian diffusion process is assumed for simplicity.
        
        The term $-\lambda_O O$ accounts for physiological oxygen consumption by healthy tissue at reference density.

        Oxygen consumption by tumor cells is modeled through the nonlinear term $-\zeta_O \rho O$.
        The consumption rate is assumed phenotype-independent for simplicity, as the present framework does not explicitly resolve alternative metabolic pathways or phenotype-dependent oxygen-consumption mechanisms.
        
        Since the previous term already accounts for oxygen consumption by healthy tissue at physiological density, the tumor contribution is interpreted as an effective excess uptake associated with the replacement of healthy cells by tumor cells.
                
        The oxygen variable \(O(t,x)\) is expressed throughout the paper in oxygen partial-pressure units (mmHg). Parameter values for $\lambda_O$ and $\zeta_O$ are reported in Table~\ref{tab:parameters}, and further details on their calibration are provided in Supporting Information S6. These parameters are chosen to ensure that oxygen dynamics rapidly reach quasi-steady conditions relative to the slower timescale of cell population dynamics.
        
        Finally, the source term $V_O(x)$ represents oxygen supply from the vasculature.  
        Allowing $V_O$ to depend on space enables the modeling of heterogeneous vascular structures, ranging from simplified and controlled configurations (e.g., \textit{in vitro} conditions) to more complex and spatially irregular oxygen inputs characteristic of \textit{in vivo} tumor environments.
        In this work, \(V_O\) is assumed to be time-independent for the sake of model simplicity, since vascular dynamics are not explicitly modeled and lie outside the scope of the present study.

\section{Results}
    We use the model to compare radiotherapy schedules across oxygenation regimes, using time-to-progression as the primary outcome. We first analyze spatially uniform oxygen supply to characterize the dose--interval response landscape and its mechanistic drivers, and then examine how toxicity constraints, intermediate oxygenation levels, and spatial oxygen-source heterogeneity reshape protocol ranking.
    \label{sec:results}        
    
    \subsection{Schedule ranking under uniform oxygenation}
        \label{sec:uniform oxygen source}
        A natural starting point for applying the model is to consider tumor evolution under a spatially uniform oxygen supply.
        In the absence of tumor cells ($\rho=0$) and under spatially uniform conditions, the steady state for the oxygen equation \eqref{eq:oxygen_dynamics} is given by $O^* = V_O/\lambda_O$. In the following, we fix the tumor-free steady state to a reference oxygenation level $I_0$, with $I_0 \in \{O_M, O_m, O_h\}$ (see Table~\ref{tab:parameters}), which determines the oxygen supply as $V_O(x) = I_0 \lambda_O$ for all $x \in \Omega_s$, yielding a homogeneous and constant oxygen input.
        From a biological perspective, this setting can be interpreted as tumor evolution occurring near a vascular structure, such as a single vessel or a cluster of vessels, resulting in an approximately uniform oxygen supply along the spatial direction under consideration and constant over time.
        Under this configuration, oxygen rapidly becomes nearly uniform within the tumor core, with only small boundary-layer deviations, owing to its fast diffusion relative to the timescale of cell dynamics.
        For this reason, this scenario serves as a reference configuration before considering more complex spatial oxygen distributions.

        Normal tissues differ in their sensitivity to fraction size, which in the LQ framework is characterized by the $(\alpha/\beta)_H$ ratio. 
        Protocol comparisons are therefore performed under a fixed constraint on the biologically effective dose (BED), defined in Supporting Information S2, by requiring that the BED of each admissible protocol does not exceed that of a conventional standard-of-care (SoC) schedule 
        $\mathcal{P}_{SoC}=(d_{SoC}=1.8~\mathrm{Gy}, \Delta_{SoC}=1~\mathrm{day}, N_{f,SoC}=30)$, namely
        \begin{equation}
        \mathrm{BED}(\mathcal{P}) \leq \mathrm{BED}(\mathcal{P}_{SoC}) .
        \end{equation}
        For each prescribed dose per fraction $d$, the number of fractions is chosen as the largest integer compatible with this constraint,
        \begin{equation}
            N_f=
            \left\lfloor
            \frac{\mathrm{BED}(\mathcal{P}_{SoC})}
            {d\left(1+d/(\alpha/\beta)_H\right)}
            \right\rfloor .
        \end{equation}
        Equivalently, among all integer numbers of fractions satisfying the BED constraint, we select the one that yields the largest BED not exceeding $\mathrm{BED}(\mathcal{P}_{SoC})$. 
        Thus, for each dose--interval pair $(d,\Delta)$, the protocol $\mathcal{P}=(d,\Delta,N_f)$ is uniquely specified. 
        This construction enforces comparable levels of healthy-tissue toxicity across treatment schedules. By preventing an artificial increase in the effective dose, it allows the model to test the effects of fractionation and temporal scheduling separately from trivial dose escalation.
            
        Following the setup proposed by \citet{HenaresMolina2017}, we explore alternative radiotherapy schedules by varying the dose per fraction $d$ and the inter-fraction interval $\Delta$. Calendar effects associated with clinical delivery schedules, such as weekend breaks, are not explicitly considered in the present model.
        The explored dose range is $d \in (0,5]\,\mathrm{Gy}$, consistent with the validity limits of the LQ model, which is known to lose accuracy beyond this interval \citep{Kirkpatrick2008, Cui2022}. Time intervals are taken as $\Delta \in (0,100]$ days, again following \citet{HenaresMolina2017}.
        This extended interval range is introduced to reveal the structural dependence of treatment outcome on the dose--interval trade-off. Although many of the explored intervals exceed typical clinical practice, they are included to characterize the underlying response behavior. As shown below, slices of the treatment landscape along either the dose or interval axis retain a similar qualitative structure, with systematic shifts in response, indicating a robust correlation between fraction spacing and dose rather than isolated optima.
        
\begin{figure*}[h]
            \centering
            \begin{subfigure}[t]{0.495\textwidth}
                \centering
                \includegraphics[width=\textwidth]{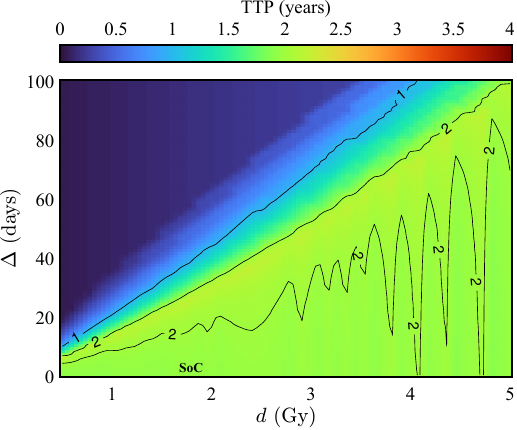}
            \end{subfigure}
            \hfill
            \begin{subfigure}[t]{0.495\textwidth}
                \centering
                \includegraphics[width=\textwidth]{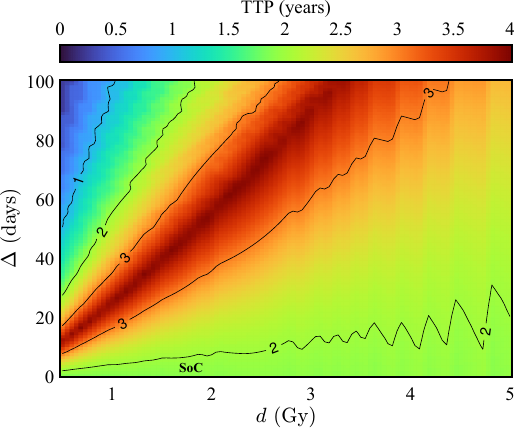}
            \end{subfigure}
            \captionsetup{
                font=small,
            }
            \caption{
            Time-to-progression (TTP), measured in years, as a function of single-fraction
            dose and inter-fraction interval under a spatially uniform oxygen supply.
            \textbf{Left:} normoxic conditions (\(I_0=O_M\)).
            \textbf{Right:} hypoxic conditions (\(I_0=O_m\)).
            Protocol comparisons are performed under the constraint
            \(\mathrm{BED}(\mathcal{P}) \leq \mathrm{BED}(\mathcal{P}_{\mathrm{SoC}})
            \approx 64\,\mathrm{Gy}\), where the standard-of-care protocol
            \(\mathcal{P}_{\mathrm{SoC}}\) consists of \(30\) fractions of
            \(1.8\,\mathrm{Gy}\) delivered at \(1\)-day intervals.
            Healthy-tissue fractionation sensitivity is characterized by \((\alpha/\beta)_H = 10\,\mathrm{Gy}\),
            consistent with values reported in \citet{HenaresMolina2017}. 
            Oxygenation qualitatively reshapes the dose--interval response landscape: under normoxia, protocol performance is relatively insensitive to fractionation choice, whereas under hypoxia, protracted schedules with longer inter-fraction intervals achieve substantially greater TTP than the standard-of-care schedule.
            }           
            \label{Fig.:TTP_maps}
        \end{figure*}

        Fig.~\ref{Fig.:TTP_maps} displays the predicted time-to-progression (TTP) across the $(d,\Delta)$ parameter space under spatially uniform oxygen supply, in normoxic (left panel) and hypoxic (right panel) conditions. Here, TTP denotes the time elapsed from the initiation of radiotherapy to tumor relapse, defined as the time at which the total tumor burden returns to the prescribed detection threshold; its formal definition is reported in Supporting Information S2. Each dose--interval map was obtained through a systematic scan of the protocol space based on \(2500\) independent model runs. Each run corresponds to a single dose--interval pair $(d,\Delta)$, for which the number of fractions is determined by the BED constraint and the resulting TTP is recorded.
        
        Under high oxygenation, the TTP landscape is relatively flat, with a broad region of dose--interval combinations yielding comparable outcomes. In this regime, the predicted time-to-progression remains close to the value obtained under the SoC schedule, typically around two years, and deviations from this reference produce only marginal variations in relapse timing.

        In contrast, under hypoxic conditions the structure of the map changes qualitatively. A distinct region of the parameter space emerges in which TTP values increase substantially, reaching almost four years along a well-defined ridge in the $(d,\Delta)$ plane. This structure delineates a \emph{therapeutic efficiency frontier}, separating protocols with limited benefit from schedules capable of nearly doubling the time-to-progression relative to standard fractionation. Notably, the frontier is located in the region of the map characterized by longer inter-fraction intervals, indicating that more protracted or metronomic-like fractionation schedules can achieve the largest TTP gains under hypoxic conditions. 
        The linear-like geometry of this high-TTP region further indicates the presence of a persistent trade-off between dose per fraction and inter-fraction spacing: rather than a single isolated optimum, comparable outcomes can be achieved along a continuous band of dose--interval combinations in the $(d,\Delta)$ plane.

    \subsection{Interplay between reoxygenation and phenotypic selection}
        \label{sec:mechanistic_interpretation}
        To qualitatively elucidate the mechanisms underlying the observed structure of the dose--interval map, we compare tumor dynamics generated by two representative protocols located in distinct regions of Fig.~\ref{Fig.:TTP_maps}: the SoC schedule and a protocol attaining maximal TTP within the same map. These cases serve as prototypical examples of low- and high-performance regimes and illustrate the dynamical mechanisms shaping treatment response. The resulting insights are consistent with the broader analysis presented in Section~\ref{sec:uniform_intermediate_levels}, where we systematically explore a range of oxygenation levels and evaluate the distribution of gains across near-optimal protocols. This analysis reveals a non-monotone dependence of treatment benefit on oxygen availability, with intermediate hypoxia emerging as the regime in which fractionation and timing have the largest impact.
        
\begin{figure*}[h]
            \centering
            \begin{subfigure}[t]{0.433\textwidth}
                \centering
                \includegraphics[width=\textwidth]{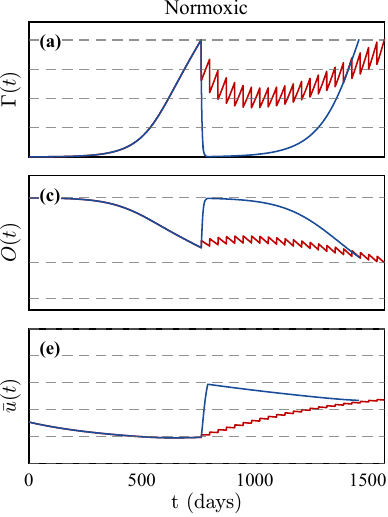}
            \end{subfigure}
            \hfill
            \begin{subfigure}[t]{0.554\textwidth}
                \centering
                \includegraphics[width=\textwidth]{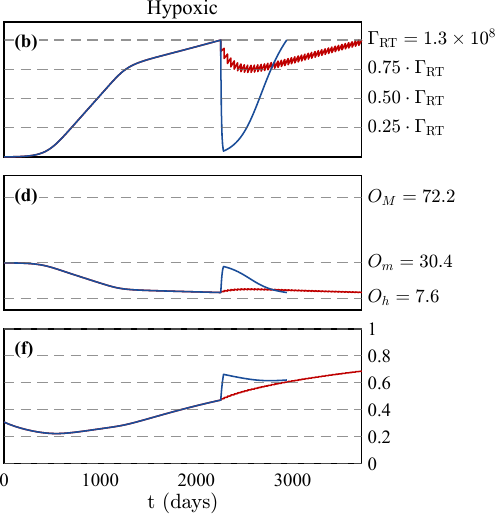}
            \end{subfigure}
            \captionsetup{font=small}
\caption{
Time evolution of tumor burden (top), oxygen level (middle), and mean phenotypic trait (bottom) under the standard-of-care protocol (blue solid line) and the maximal-TTP protocol (red solid line).Panels (a,b) show tumor burden, panels (c,d) the oxygen level, and panels (e,f) the mean phenotypic trait. Tumor burden \(\Gamma(t)\) is expressed as the total number of cells, while oxygen levels and thresholds are expressed in \(\mathrm{mmHg}\).
The standard-of-care protocol consists of \(30\) fractions of \(1.8\,\mathrm{Gy}\) delivered at \(1\)-day intervals.
The detection threshold \(\Gamma_{\mathrm{RT}}\) and the oxygenation thresholds \(O_M\), \(O_m\), and \(O_h\) are reported in Table~\ref{tab:parameters}.
Panels (a,c,e) correspond to the normoxic case \((I_0 = O_M)\): the maximal-TTP protocol delivers approximately \(2.3\,\mathrm{Gy}\) every \(15\) days for \(22\) fractions (total dose \(\sim 51\,\mathrm{Gy}\)), yielding a TTP gain of approximately \(113\) days.
Panels (b,d,f) correspond to the hypoxic case \((I_0 = O_m)\): the maximal-TTP protocol delivers approximately \(1.3\,\mathrm{Gy}\) every \(35\) days for \(42\) fractions (total dose \(\sim 55\,\mathrm{Gy}\)), yielding a TTP gain of approximately \(2\) years. Treatment-induced reoxygenation interacts differently with phenotypic selection across oxygenation regimes: under normoxia, SoC-induced reoxygenation is partially offset by redistribution toward resistant phenotypes, whereas under hypoxia resistance is already established, leading to earlier relapse under SoC than under the protracted maximal-TTP schedule.
}
            %2.336 Gy 37.051 days 
            % 1.326 Gy 35.02 days
            \label{Fig.:Comparison_frontiers}
        \end{figure*}

        Fig. \ref{Fig.:Comparison_frontiers} shows the time evolution of tumor burden, oxygen concentration, and mean phenotype (see Supporting Information S2 for its definition) for both protocols under normoxic and physiological hypoxic conditions.
        During the initial stage, before treatment delivery, the time evolution of the three reported quantities coincides for the two protocols, since tumor growth and the associated microenvironmental pressure are identical. Differences emerge only after radiation delivery, when the trajectories associated with the two protocols start to diverge.
        In the hypoxic case  (panels (b,d,f) of Fig.~\ref{Fig.:Comparison_frontiers}), the visible change in slope of the pre-treatment tumor-burden curve around \(t\approx1000\) days reflects a progressive slowing of tumor growth as the population expands. Increasing local cell density strengthens the crowding contribution in Eq.~\eqref{eq:reaction_term}, while the accompanying decrease in oxygen availability further limits proliferation and promotes selection toward more hypoxia-adapted, less proliferative phenotypes. These combined effects reduce the net growth rate and explain the longer time required to reach the detection threshold \(\Gamma_{\mathrm{RT}}\) compared with the normoxic case.

        As shown in panels (a,c,e) of Fig.~\ref{Fig.:Comparison_frontiers}, under normoxic conditions, the maximal-TTP protocol operates within a mildly deoxygenated TME and induces only a limited phenotypic shift. In contrast, the SoC protocol promotes tumor reoxygenation while delivering more densely spaced radiation fractions, thereby redistributing the population toward higher-resistance trait values. As a result, microenvironmental effects and therapy-induced selection partially offset one another, leading to comparable TTP values despite markedly different underlying dynamics.
        
        Under hypoxic conditions (panels (b,d,f) of Fig.~\ref{Fig.:Comparison_frontiers}), the maximal-TTP protocol again maintains low oxygenation levels with only minor changes in the phenotypic distribution. The SoC protocol still increases oxygen availability; however, unlike the normoxic case, this does not induce a substantial redistribution of phenotypes. In this regime, the TME has already selected for highly resistant phenotypes, leaving little room for further therapy-driven phenotypic switching. Consequently, the balance observed under normoxia disappears: resistance is already established, resulting in earlier relapse of the SoC protocol after reoxygenation.
        
    \subsection{Impact of toxicity constraints}
        \label{sec:toxicity_constraints}
        So far, protocol comparisons have been performed under the constraint that the equivalent toxicity does not exceed that of a SoC schedule. This assumption, however, is not universally appropriate. In some clinical scenarios, dose escalation may be considered to improve tumor control \citep{Brower2016}, whereas in others stricter toxicity limits may require a reduction of the admissible dose \citep{Timmerman2006}.
        
        From a clinical standpoint, treatment planning is inherently multi-objective, involving a trade-off between tumor control probability (TCP) \citep{Hillen2013} and normal-tissue toxicity, commonly quantified through the Normal Tissue Complication Probability (NTCP) \citep{Gaito2022}. Additional variability arises from patient- and site-specific factors, including intrinsic radiosensitivity \citep{Nuijens2025} and the geometric configuration of organs at risk (OARs) \citep{Eber2025}. These considerations suggest that differences in normal-tissue responsiveness and tolerance to fraction size modulate the feasible treatment space and should therefore be incorporated into protocol comparison.
        
        The aim of the following analysis is to characterize how protocol efficiency depends on oxygen availability, normal-tissue radiosensitivity, and admissible toxicity levels.
        To this end, we extend the exploration of simulated radiotherapy schedules by considering multiple combinations of BED, normal-tissue radiosensitivity $(\alpha/\beta)_H$, and oxygen availability.
        
        TTP is again adopted as the efficacy metric. The therapeutic efficiency frontier -- formally defined in Supporting Information S7 -- is identified as the subset of dose--interval combinations attaining at least $80\%$ of the maximum predicted TTP.
        For each parameter configuration, we compute a TTP-weighted barycenter of the frontier (see Supporting Information S7 for details). This barycenter is not intended to identify an optimal schedule; rather, it provides a compact descriptor of the region in the dose--interval plane associated with highly effective regimens, thereby enabling systematic comparison across oxygenation and toxicity scenarios.
        
        Nine parameter configurations are analyzed by combining three oxygenation states---normoxic $(I_0 = O_M)$, hypoxic $(I_0 = O_m)$, and severely hypoxic $(I_0 = O_h)$---with three paired toxicity settings: $(\mathrm{BED},(\alpha/\beta)_H)=(90\,\mathrm{Gy},10\,\mathrm{Gy})$, $(60\,\mathrm{Gy},6\,\mathrm{Gy})$, and $(30\,\mathrm{Gy},2\,\mathrm{Gy})$.
        
\begin{figure*}[h!]
            \centering    
            % --- (a) ---
            {\small\textbf{(a)} Normoxic conditions ($I_0 = O_M$)}\\[0.5ex]
            \includegraphics[width=0.32\textwidth]{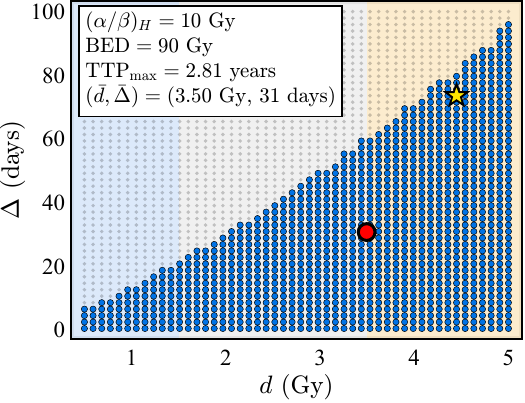}
            \includegraphics[width=0.32\textwidth]{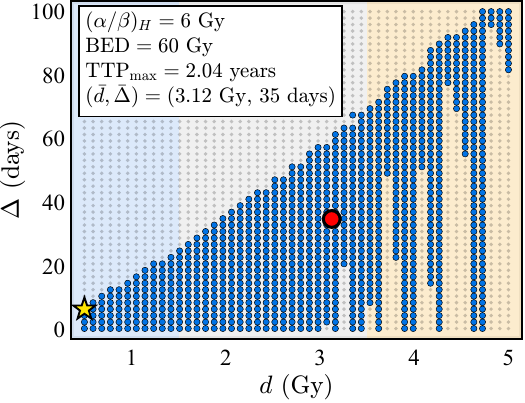}
            \includegraphics[width=0.32\textwidth]{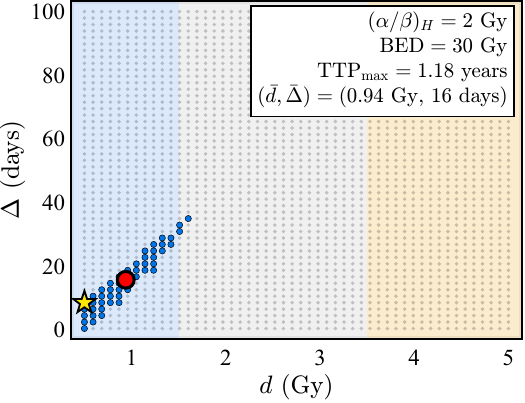}
            
            \vspace{1.5ex}
            
            % --- (b) ---
            {\small\textbf{(b)} Mildly hypoxic conditions ($I_0 = O_m$)}\\[0.5ex]
            \includegraphics[width=0.32\textwidth]{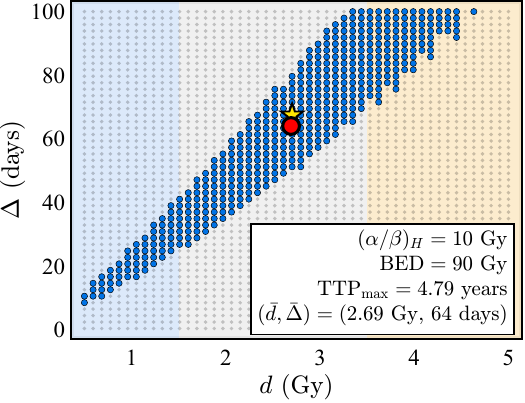}
            \includegraphics[width=0.32\textwidth]{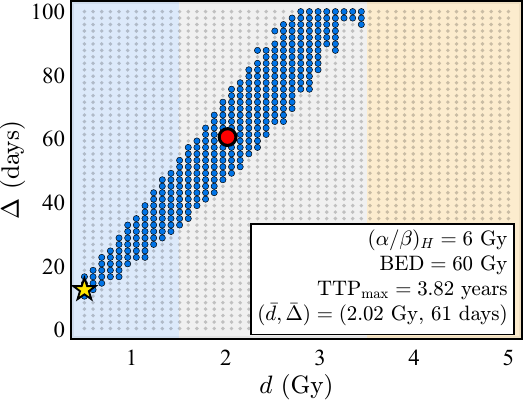}
            \includegraphics[width=0.32\textwidth]{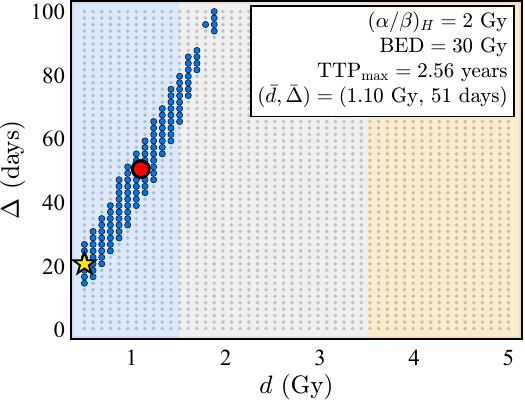}
            
            \vspace{1.5ex}
            
            % --- (c) ---
            {\small\textbf{(c)} Severely hypoxic conditions ($I_0 = O_h$)}\\[0.5ex]
            \includegraphics[width=0.32\textwidth]{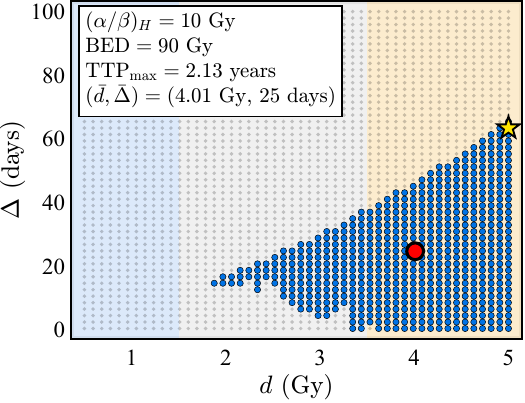}
            \includegraphics[width=0.32\textwidth]{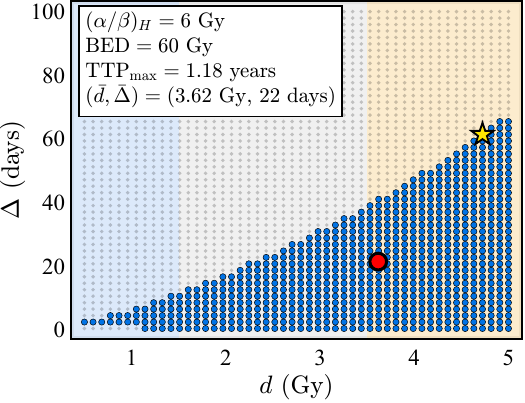}
            \includegraphics[width=0.32\textwidth]{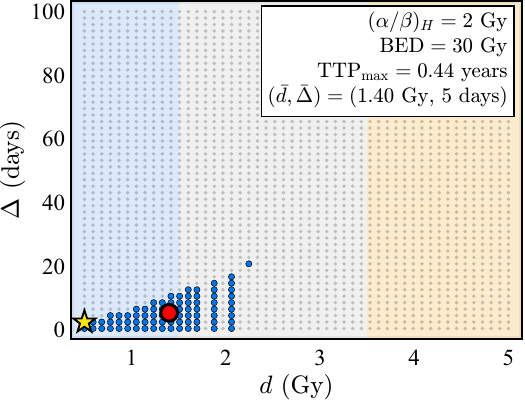}
            
            \captionsetup{font=small}
            \caption{
            Therapeutic efficiency frontiers in the dose--interval plane under spatially uniform oxygen supply.
            Each column corresponds to a fixed normal-tissue sensitivity \((\alpha/\beta)_H\) and BED constraint, common to all oxygenation regimes shown in that column.
            Rows correspond to increasing hypoxia:
            \textbf{(a)} normoxic conditions \((I_0=O_M)\),
            \textbf{(b)} moderately hypoxic conditions \((I_0=O_m)\), and
            \textbf{(c)} severely hypoxic conditions \((I_0=O_h)\).
            For each parameter configuration, the therapeutic efficiency frontier contains the protocol combinations achieving at least \(80\%\) of the maximum TTP.
            Star markers indicate the dose--interval pair achieving maximal TTP, while circular markers indicate the corresponding TTP-weighted barycenters.
            The dose--interval plane is partitioned into three regions to distinguish schedules typically classified as hyperfractionated (\(d \leq 1.5\,\mathrm{Gy}\)), standard-to-moderately hypofractionated (\(1.5\,\mathrm{Gy}< d < 3.5\,\mathrm{Gy}\)), and strongly hypofractionated (\(3.5\,\mathrm{Gy} \leq d \leq 5\,\mathrm{Gy}\)). Oxygenation and normal-tissue tolerance jointly reshape the therapeutic efficiency frontier. Protracted schedules are favored under moderate hypoxia, whereas under severe hypoxia hypofractionated schedules can be advantageous at higher tolerance levels.
}            
            \label{Fig.:Frontiers_Comparison}
        \end{figure*}

        Under normoxic conditions (Fig.  \ref{Fig.:Frontiers_Comparison}a), the therapeutic efficiency frontier spans a broad region of the dose--interval plane, reflecting the limited sensitivity of TTP to protocol variations in this regime. A wide range of schedules yields comparable outcomes, consistent with the relatively flat TTP landscape observed under uniform oxygenation. As toxicity constraints become more restrictive, the admissible region narrows and the frontier barycenter shifts toward more hyperfractionated schedules.
                         
        As oxygen availability decreases, the therapeutic efficiency frontier exhibits a clear diagonal structure (Fig. \ref{Fig.:Frontiers_Comparison}b). The maximal predicted TTP also increases substantially relative to normoxic conditions, in some cases approximately doubling across the considered toxicity constraints. In this regime, longer inter-fraction intervals are favored, consistently with the mechanism discussed in Section~\ref{sec:mechanistic_interpretation}: hypoxia-driven selection stabilizes slowly proliferating, hypoxia-adapted resistant phenotypes, weakens the compensatory effect of reoxygenation, and thereby reduces the benefit of closely spaced irradiation. Reduced normal-tissue tolerance narrows the diagonal and increases its slope. Since higher doses are no longer admissible, compensation occurs through longer inter-fraction intervals. In this configuration, the frontier is largely confined to protracted schedules over a broad range of admissible toxicity levels. The barycenter therefore lies in the metronomic region, indicating that such schedules remain robust across varying tissue tolerance conditions.
                 
        In severely hypoxic conditions (Fig. \ref{Fig.:Frontiers_Comparison}c), the therapeutic efficiency frontier no longer exhibits the regular diagonal structure observed at higher oxygen levels. In addition, the maximal achievable TTP decreases relative to the mildly hypoxic regime, reflecting the reduced effectiveness of radiation under extreme oxygen deprivation. In this setting protocol selection is primarily governed by normal-tissue tolerance. When tolerance is relatively high, hypofractionated schedules dominate the admissible region, whereas decreasing tolerance induces a transition toward hyperfractionated schedules, which become progressively accessible. Consistently, the barycenter shifts from hypofractionated to hyperfractionated regions of the dose--interval plane as toxicity constraints become more restrictive. This observation aligns with the dynamics of slowly proliferating, late-responding tumors and is consistent with clinical evidence showing that tumors such as prostate cancer may benefit from hypofractionated treatment schedules \citep{Fowler2001}. 

        This result provides a complementary perspective to standard radiobiological modeling by incorporating phenotypic structure. Classical LQ--OER formulations, which are largely empirical and do not account for phenotypic heterogeneity, typically associate hypoxia with an increased $\alpha/\beta$ ratio, thereby reducing the predicted benefit of hypofractionation. In contrast, our framework couples oxygenation with phenotypic selection, so that the $\alpha/\beta$ ratio emerges as a dynamic quantity. As detailed in Supporting Information S5, this mechanism can lead to a decrease of the $\alpha/\beta$ ratio under hypoxia, consistent with clinical observations and enabling the emergence of hypofractionation-favorable regimes.

    \subsection{Extension to intermediate oxygenation levels}
        \label{sec:uniform_intermediate_levels}
        
        We next refine the uniform-oxygenation analysis by considering a denser set of oxygen supply levels between pathological hypoxia and normoxia, \(I_0\in[O_h,O_M]\). For each value of \(I_0\), we reconstructed the dose--interval TTP map and selected the corresponding therapeutic efficiency frontier, using the same definition as in Supporting Information S7. In particular, the frontier identifies the subset of dose--interval combinations that achieve near-maximal TTP values under the imposed constraints, thereby capturing the region of high-performing treatment protocols. In this analysis, all protocols were generated under the BED constraint associated with the standard-of-care schedule \(\mathcal{P}_{\mathrm{SoC}}\), with healthy-tissue fractionation sensitivity \((\alpha/\beta)_H=10\) Gy, as introduced in Section~\ref{sec:uniform oxygen source}.
        
\begin{figure}
            \centering
            \includegraphics[width=\linewidth]{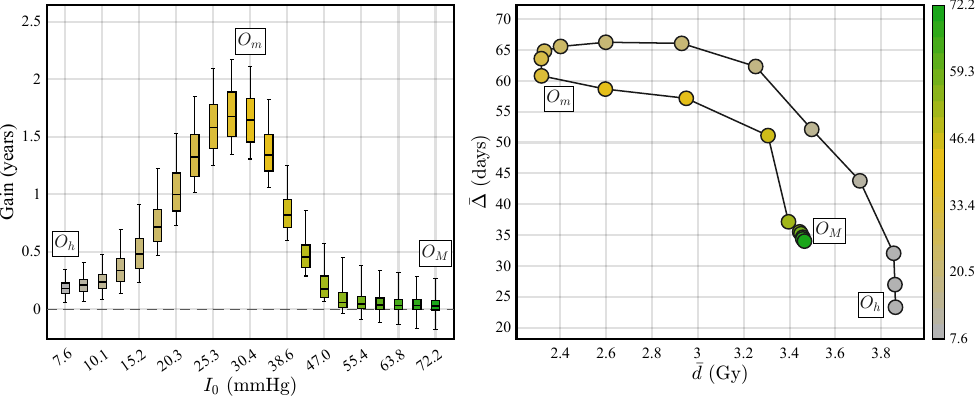}
            \captionsetup{font=small}
            \caption{
            Extension of the uniform oxygenation analysis to intermediate oxygen supply levels. 
            For each value of \(I_0\), the dose--interval TTP map is reconstructed from \(2500\) independent model runs, and the therapeutic efficiency frontier is then extracted and analyzed.
            \textbf{Left:} distribution of TTP gains, measured relative to the standard-of-care protocol, over the corresponding efficiency frontier. Boxes indicate the interquartile range, central lines denote medians, and whiskers indicate minimum and maximum gains. Colors encode oxygenation level. The dashed line marks zero gain. \textbf{Right:} TTP-weighted barycenter of the same efficiency frontier in the dose--interval plane as a function of \(I_0\). 
            The benefit of departing from standard-of-care fractionation is non-monotone in oxygen availability, peaking at intermediate hypoxia and becoming limited under both normoxia and severe hypoxia.            
            }
            \label{Fig.:uniform_livelli}
        \end{figure}

        For each oxygenation level, let \(\mathcal{F}(I_0)\) denote the set of protocols belonging to the therapeutic efficiency frontier. We then evaluate how these near-optimal protocols perform relative to the standard-of-care schedule under the same oxygenation condition. To this end, for every protocol \(\mathcal{P}\in\mathcal{F}(I_0)\), we define the TTP gain as
        \begin{equation}
        g_{\mathcal{P}}(I_0)
        =
        \mathrm{TTP}_{\mathcal{P}}(I_0)
        -
        \mathrm{TTP}_{\mathrm{SoC}}(I_0).
        \end{equation}
        The distribution of \(g_{\mathcal{P}}(I_0)\) over \(\mathcal{F}(I_0)\) was summarized by its median, minimum and maximum values, together with lower and upper quartile indicators. These quantities were used to construct the boxplot-like representation shown in Fig.~\ref{Fig.:uniform_livelli} (left panel). 
        
        This analysis shows that the benefit of alternative fractionation schedules is not a monotone function of oxygen availability. In well-oxygenated conditions, the gain is negligible or only marginal, consistently with the relatively flat TTP landscape observed under normoxia. As oxygenation decreases, the gain increases and reaches its largest values in the intermediate hypoxic range. This identifies a window in which the standard-of-care schedule is no longer close to the most effective region of the dose--interval landscape. Under severe hypoxia, however, the gain decreases again. This non-monotone behavior suggests that moderate hypoxia is the regime in which fractionation and timing have the largest impact: oxygen is sufficiently low to reshape phenotypic selection and protocol ranking, but not so low that radiation efficacy is globally suppressed. This observation is consistent with the mechanistic picture outlined in Section \ref{sec:mechanistic_interpretation}, where the interplay between reoxygenation and phenotypic selection governs treatment response across oxygenation regimes.
    
        Using the TTP-weighted barycenter introduced above as a compact descriptor of each frontier, Fig.~\ref{Fig.:uniform_livelli} (right panel) provides a roadmap of how the high-performance region of the dose--interval plane shifts with \(I_0\). In normoxic conditions, the frontier is centered on moderately hypofractionated schedules. As oxygenation decreases toward moderate hypoxia, the barycenter moves toward lower doses per fraction and longer inter-fraction intervals, corresponding to more protracted, metronomic-like schedules. In severe hypoxia, the trajectory bends back toward larger doses per fraction and shorter intervals, consistently with the behavior observed under the previous toxicity-constraint analysis.

        Taken together, the two panels of Fig.~\ref{Fig.:uniform_livelli} provide a consistent picture. The non-monotone shift of the frontier across oxygenation levels (right panel) is accompanied by a corresponding non-monotone trend in the gain distribution (left panel), with maximal improvements observed in the intermediate hypoxic range. In normoxia, gains remain limited, whereas in severe hypoxia the reduction in oxygen-mediated radiosensitization constrains the achievable benefit despite the shift toward larger fraction sizes. Overall, these results identify intermediate hypoxia as the regime in which deviations from standard fractionation are most consequential, with a systematic shift toward more protracted schedules.
    
    \subsection{Role of oxygen spatial heterogeneity}
    \label{sec:heterogeneous_case}

        The assumption of spatially uniform supply neglects the intrinsic geometric heterogeneity of tumor vasculature. Spatial variations in oxygen concentration are expected to modify local radiosensitivity, reshape eco-evolutionary tumor dynamics, and ultimately alter protocol ranking.
        
         To isolate this geometric effect, we shift our focus from the uniform configuration to an ensemble of spatially heterogeneous oxygen sources associated with the same reference tumor-free oxygenation level $I_0$. In the following, we consider two representative oxygenation regimes, corresponding to normoxic ($I_0 = O_M$) and moderately hypoxic ($I_0 = O_m$) conditions.
        
        For each of these regimes, we generate multiple realizations of the oxygen source, each corresponding to a different spatial redistribution of the same total amount. In practice, each realization is obtained by combining a finite number of localized contributions with randomly sampled positions, amplitudes, and spatial extents, and then rescaled so as to preserve the same global oxygen input. This construction produces a broad range of spatial configurations, from nearly uniform profiles to strongly localized and polarized distributions. As a result, differences in treatment outcome can be attributed to the geometry of oxygen delivery rather than to changes in the overall oxygen supply. Details on the construction of these heterogeneous oxygen sources are provided in Supporting Information S8.
        
        Under these conditions, identical radiotherapy schedules may produce substantially different times-to-progression, highlighting the role of microenvironmental geometry in determining relapse dynamics.

        We then assess the robustness of protocol selection with respect to this spatial
        perturbation. The analysis is designed to separate two distinct steps: protocol
        selection under a simplified homogeneous assumption, and protocol evaluation
        under heterogeneous oxygen delivery. The procedure is applied independently to
        the two reference oxygenation regimes considered in this section,
        \(I_0\in\{O_M,O_m\}\).
        
        For each value of \(I_0\), we first identify the therapeutic efficiency frontier
        \(\mathcal{F}_u(I_0)\) under spatially uniform oxygen supply, constructed
        exactly as in Section~\ref{sec:uniform_intermediate_levels}. Thus,
        \(\mathcal{F}_u(I_0)\) is the set of protocols that would be selected as
        near-optimal if only the corresponding uniform oxygen input were available.
        
        We next test these same protocols on heterogeneous oxygen-source
        configurations, without redefining the frontier and without re-optimizing the
        schedules for each geometry. Specifically, for each value of \(I_0\), we
        generate \(M=100\) heterogeneous oxygen-source Monte Carlo realizations
        \begin{equation}
            V_{O,I_0}^{(m)}(x),
            \qquad m=1,\ldots,M,
        \end{equation}
        all normalized to have the same total oxygen input as the corresponding
        uniform-source case. For each realization, we evaluate the same dose--interval
        grid used in the uniform analysis, consisting of \(2500\) protocols satisfying
        the BED constraint. Therefore, the heterogeneous-source analysis consists of
        \(2500\times 100\) simulations per oxygenation regime, and
        \(2500\times 100\times 2\) simulations in total across the two regimes. These
        heterogeneous simulations are used to evaluate the performance of protocols
        selected from \(\mathcal{F}_u(I_0)\), rather than to construct a new
        geometry-specific efficiency frontier.
        
        For each value of \(I_0\), each heterogeneous realization \(m\), and each
        protocol \(\mathcal{P}\in\mathcal{F}_u(I_0)\), we compute the gain
        \begin{equation}
        g_{\mathcal{P}}^{(m)}(I_0)
        =
        \mathrm{TTP}_{\mathcal{P}}^{(m)}(I_0)
        -
        \mathrm{TTP}_{\mathrm{SoC}}^{(m)}(I_0),
        \end{equation}
        where both terms are evaluated under the same oxygen-source realization
        \(V_{O,I_0}^{(m)}(x)\). This approach compares each protocol with the
        standard-of-care schedule within the same oxygenation regime and spatial
        configuration, so that the resulting gain measures the benefit of protocol
        choice rather than the absolute effect of a more or less favorable geometry.
        
        To quantify the degree of spatial heterogeneity, each realization
        \(V_{O,I_0}^{(m)}(x)\) is assigned a \emph{Gini coefficient} \(\mathcal{G}_{I_0}^{(m)}\), computed from the corresponding discretized
        oxygen-source profile. This index primarily captures the spatial polarization
        of oxygen supply, with zero Gini coefficient corresponding to the perfectly
        uniform source and larger values indicating increasingly localized oxygen
        delivery. We use the Gini coefficient as a compact descriptor of heterogeneity, although other spatial indicators could in principle capture complementary geometric features.
        
\begin{figure}[h]
            \centering
            \includegraphics[width=\linewidth]{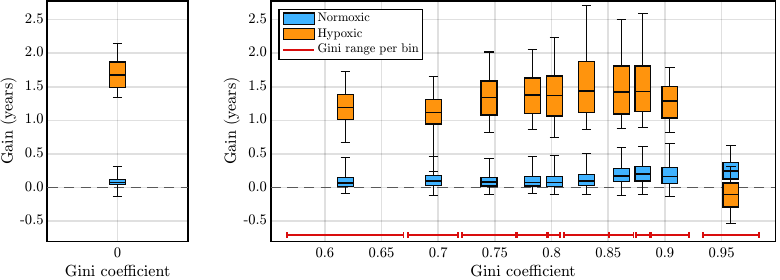}
            \captionsetup{font=small}
            \caption{
            Effect of spatial oxygen-source heterogeneity on protocols selected under
            uniform-supply assumptions. For each \(I_0\in\{O_M,O_m\}\), protocols belonging
            to the uniform-source frontier \(\mathcal{F}_u(I_0)\) are evaluated under either
            the corresponding uniform source or heterogeneous oxygen-source configurations,
            without re-optimization. Gains are measured relative to the standard-of-care
            protocol under the same oxygenation regime and spatial configuration.
            \textbf{Left:} uniform-source reference cases, for which the Gini coefficient is
            zero. \textbf{Right:} heterogeneous configurations ordered by increasing Gini
            coefficient and grouped into bins of 10 realizations, separately for each
            \(I_0\). Blue boxes indicate normoxic conditions (\(I_0=O_M\)); orange boxes
            indicate moderately hypoxic conditions (\(I_0=O_m\)). Boxes represent the
            interquartile range, central lines denote medians, and whiskers indicate minimum
            and maximum gains. Red segments indicate the Gini range of each bin, and the
            dashed line marks zero gain.
            Protocols selected as near-optimal under spatially uniform oxygenation remain qualitatively robust under heterogeneous oxygen delivery, but their individual benefit becomes increasingly variable and can turn negative as spatial heterogeneity increases, especially under hypoxia.
            }
            \label{Fig.:Gini}
        \end{figure}

        Fig.~\ref{Fig.:Gini} is constructed as follows. The left panel reports the
        uniform-source reference cases. For each oxygenation regime, the corresponding
        boxplot summarizes the distribution of gains obtained by protocols in
        \(\mathcal{F}_u(I_0)\) under spatially uniform oxygen supply. The right panel
        reports the heterogeneous cases. For each value of \(I_0\), the geometries
        are ordered by increasing \(\mathcal{G}_{I_0}^{(m)}\) and grouped into
        consecutive bins of 10 geometries each. For each bin, the boxplot summarizes
        the pooled set of gains
        \begin{equation}
            \left\{
            g_{\mathcal{P}}^{(m)}(I_0)
            :
            \mathcal{P}\in\mathcal{F}_u(I_0),\
            m\in B_k(I_0)
            \right\},
        \end{equation}
        where \(B_k(I_0)\) denotes the \(k\)-th Gini bin for the oxygenation regime
        \(I_0\). Boxes indicate the interquartile range, central lines denote medians,
        and whiskers indicate minimum and maximum gains. The horizontal red segment
        associated with each bin reports the corresponding Gini interval, from the
        minimum to the maximum \(\mathcal{G}_{I_0}^{(m)}\) among the geometries in that
        bin. Thus, each box describes how the gain of protocols selected under uniform
        oxygenation varies across heterogeneous geometries with comparable Gini
        coefficient. The two series of boxplots correspond to the two total oxygen
        supplies \(I_0=O_M\) and \(I_0=O_m\). Several features emerge.
                
        Spatial heterogeneity induces a marked increase in variability. Even when restricted to protocols belonging to the efficiency frontier under uniform conditions, the achieved gains span a wide range, including negative values. This indicates that, while the homogeneous frontier identifies high-performing protocols on average, their performance can vary substantially depending on the underlying spatial configuration.
        
        Moreover, the effect of heterogeneity on treatment outcome is strongly regime-dependent. In normoxic conditions, the median gain remains approximately constant, reflecting the overall flatness of the response landscape and the limited sensitivity to protocol selection. In contrast, under hypoxic conditions, spatial heterogeneity leads to both larger potential improvements and substantially increased variability. In particular, for highly heterogeneous configurations (large Gini coefficient), the gain can decrease significantly and may even become negative, indicating that protocols selected under homogeneous assumptions can be compromised by unfavorable spatial arrangements.
        
        An unbinned version of the same analysis, in which heterogeneous configurations are shown individually rather than pooled across Gini intervals, is reported in Supporting Figure S3. The same regime-dependent trend is recovered, indicating that binning is used only to improve visualization.
        
        Overall, these results indicate that the efficiency frontier identified under uniform source assumptions remains qualitatively robust under spatial redistribution of oxygen, particularly in the hypoxic regime, where protocols with longer inter-fraction intervals, for sufficiently low heterogeneity, yield substantial gains. However, this robustness is primarily qualitative: spatial heterogeneity introduces significant variability in treatment outcomes, so that the performance of a given protocol can depend strongly on the specific geometric configuration. In normoxia, the frontier continues to provide limited benefit, in agreement with the uniform case, and spatial heterogeneity does not introduce a clear advantage in protocol selection. Rather, it mainly induces variability without altering the overall flatness of the response landscape.
        These observations suggest that incorporating information on the spatial geometry of oxygen delivery could significantly improve protocol selection. In particular, geometry-aware selection may reduce the variability of TTP gains across heterogeneous configurations and, more importantly, may help exclude spatial regimes in which protocols selected under uniform assumptions yield negative gains relative to the SoC. This effect is especially relevant in hypoxic conditions, where large positive gains are possible but unfavorable geometries can compromise protocol performance. In normoxic conditions, spatial information may instead help identify the limited subset of configurations in which non-standard schedules provide a measurable advantage.
        
\section{Discussion}
    \label{sec:discussion}
       
    This study shows that spatial oxygen structure and phenotypic adaptation jointly alter radiotherapy protocol ranking. Rather than restricting the analysis to a limited number of predefined schedules under fixed radiosensitivity assumptions, we systematically explore the dose--interval plane and examine how protocol performance varies across oxygenation regimes and toxicity constraints. A summary of the main model-based findings and their implications is provided in Table~\ref{tab:key_takeaways}.

    \begin{table*}[!h]
    \centering
    \footnotesize
    \renewcommand{\arraystretch}{1.35}
    \setlength{\tabcolsep}{5pt}
    \begin{tabularx}{\textwidth}{|>{\raggedright\arraybackslash}p{0.23\textwidth}|
                                      >{\raggedright\arraybackslash}X|
                                      >{\raggedright\arraybackslash}X|}
    \hline
    \rowcolor{gray!20}
    \textbf{Model component} & \textbf{Main finding} & \textbf{Implications for protocol selection} \\
    \hline
    
    \rowcolor{takeawayBlue}
    Uniform oxygen supply &
    Under normoxic conditions, the dose--interval TTP landscape is relatively flat, whereas under moderate hypoxia a structured high-performance region emerges. &
    The ranking of fractionation schedules changes with oxygenation. In normoxia, deviations from standard fractionation provide limited benefit, while hypoxia can make treatment timing substantially more relevant. \\
    \hline
    
    \rowcolor{takeawayGreen}
    Phenotypic adaptation &
    The response to radiotherapy is shaped by the interplay between reoxygenation and selection of hypoxia-adapted, more resistant phenotypes. &
    Protocol performance depends not only on direct oxygen-mediated radiosensitivity, but also on TME-driven enrichment of slowly proliferating, hypoxia-adapted resistant phenotypes. \\
    \hline
    
    \rowcolor{takeawayYellow}
    Intermediate oxygenation levels &
    The gain over standard-of-care is non-monotone with respect to oxygen supply, reaching its largest values in the intermediate hypoxic range. &
    Moderate hypoxia appears as the regime in which schedule adaptation has the largest impact, with high-performing schedules shifting toward lower doses per fraction and longer intervals between fractions.\\
    \hline
    
    \rowcolor{takeawayBlue}
    Toxicity constraints &
    Changing normal-tissue tolerance reshapes the admissible dose--interval region and shifts the therapeutic efficiency frontier. &
    Normal-tissue tolerance can drive a switch between hypofractionated and hyperfractionated schedules, particularly under severe hypoxia, where high-tolerance settings favor larger fraction sizes whereas stricter toxicity constraints shift selection toward smaller, more frequent fractions. \\
    \hline
    
    \rowcolor{takeawayGreen}
    Spatial oxygen-source geometry &
    At fixed total oxygen supply, heterogeneous source configurations induce substantial variability in TTP gains. Schedules that are near-optimal under uniform oxygen supply remain broadly informative, but their individual performance can change substantially with oxygen-source geometry. &
    Spatial information may improve protocol selection by reducing uncertainty in hypoxic regimes and by potentially revealing geometry-dependent gains in normoxic conditions. \\
    \hline
    
    \rowcolor{takeawayYellow}
    Modeling perspective &
    Coupling oxygen dynamics with phenotypic structure provides a mechanism by which effective radiosensitivity and fractionation response can vary across oxygenation regimes. &
    Mechanistic, spatially resolved models can complement classical LQ--OER reasoning by explicitly accounting for how hypoxia-driven phenotypic selection reshapes radiosensitivity and fractionation response in heterogeneous tumors.\\
    \hline
    \end{tabularx}
    \captionsetup{font=small}
        {
    \captionsetup{font=customcaption}
    \caption{
        Key takeaways from the model analysis. The table summarizes how oxygenation regime, phenotypic adaptation, toxicity constraints, and oxygen-source geometry affect radiotherapy protocol ranking in the proposed framework.
    }
    \label{tab:key_takeaways}
    }
    \end{table*}
    
    A first central result is that the effectiveness of a fractionation schedule is not intrinsic to the schedule itself, but depends critically on the TME. Under spatially uniform oxygen supply, we identify a region of near-optimal schedules---referred to as the therapeutic efficiency frontier---defined as the subset of dose--interval combinations that achieve near-maximal time-to-progression. In practice, this corresponds to the high-performance region of the dose--interval map obtained by systematically scanning fractionation protocols. This region exhibits a relatively regular structure, and deviations from standard-of-care protocols provide limited benefit in normoxic conditions. In contrast, under hypoxia the frontier changes shape and alternative schedules---particularly protracted or metronomic-like schemes---can substantially delay tumor regrowth. At the mechanistic level, this behavior is consistent with the interplay between the well-known reoxygenation effect following radiation delivery and phenotype redistribution  \citep{Steel1989}. In particular, when reoxygenation compensates therapy-driven selection toward resistant traits, time-to-progression values remain comparable; when this compensation mechanism weakens, slower-proliferating hypoxia-adapted phenotypes alter the balance and reshape protocol performance.
    
    The structure of the therapeutic efficiency frontier is further modulated by tissue tolerance. Chan\-ges in biologically effective dose and normal-tissue $(\alpha/\beta)_H$ ratio reshape the admissible region of the dose--interval plane. Under severe hypoxia, protocol selection becomes strongly governed by normal-tissue responsiveness. When tolerance is relatively high, hypofractionated schedules dominate the frontier, whereas decreasing tolerance induces a sharp transition toward hyperfractionated schemes. This finding aligns with the dynamics of slowly proliferating, late-responding tumors and is consistent with clinical observations suggesting that selected hypoxic tumors may benefit from hypofractionated strategies~\citep{Fowler2001}.
    
    The extension to intermediate oxygen supply levels clarifies that the benefit of schedule adaptation is non-monotone with respect to oxygen availability. Gains over standard-of-care were small under normoxia, increased in the intermediate hypoxic range, and decreased again under severe hypoxia. This indicates that moderate hypoxia represents a window in which oxygen is sufficiently low to reshape phenotypic selection and protocol ranking, but not so low as to globally suppress radiation efficacy. The barycenter trajectory of the efficiency frontier supports the same interpretation: as oxygenation decreases from normoxia to moderate hypoxia, the high-performance region shifts toward more protracted schedules; under severe hypoxia, it bends back toward larger fraction sizes and shorter intervals. Thus, the model does not simply predict that 'less oxygen' uniformly favors alternative schedules. Instead, it identifies a specific oxygenation regime in which fractionation and timing have the largest effect.
    
    A further result concerns the role of spatial geometry. In the heterogeneous-source analysis, we kept the total oxygen supply fixed and varied only its spatial organization. This construction separates the effect of oxygen-source geometry from the effect of total oxygen input. Protocols selected from the uniform-source efficiency frontier were then applied to heterogeneous configurations without re-optimization. The resulting gains show that the uniform-source frontier remains qualitatively informative, especially under hypoxia, where protracted schedules often continue to provide substantial benefit. At the same time, spatial heterogeneity introduces substantial outcome uncertainty, inducing large variability in treatment response. In highly polarized configurations, the gain may decrease substantially or even become negative. Therefore, spatial structure should not be interpreted as a small perturbation of an otherwise average oxygenation state. It can affect both the expected benefit and the reliability of a protocol selected under simplified assumptions. In this perspective, spatially resolved biomarkers of hypoxia, such as those obtained from functional imaging modalities (e.g. FMISO-PET), could provide more informative descriptors of the TME and support treatment selection beyond mean oxygenation levels \citep{McNeal2024}.
    
    The model also provides a complementary perspective to standard LQ--OER formulations. Classical oxygen-dependent LQ models rescale radiosensitivity through oxygen availability but usually do not represent the phenotypic redistribution induced by hypoxic selection. Here, oxygen acts both directly, through radiosensitivity modulation, and indirectly, through the selection of phenotypes with different proliferative and radiotherapy-response traits. As discussed in the Supporting Information S5, this coupling allows the effective \(\alpha/\beta\) behavior to change along the selected phenotypic trajectory, offering a possible mechanistic bridge between controlled oxygen-response experiments and clinical observations in which hypoxic or slow-growing tumors may benefit from hypofractionated treatment.
    
    The present formulation relies on simplifying assumptions that enable a controlled investigation of treatment comparison under heterogeneous microenvironmental conditions. In particular, the spatial domain is one-dimensional, oxygen sources are prescribed rather than dynamically remodeled, and radiation-induced vascular remodeling is not explicitly represented~\citep{Zunino2020}. These assumptions limit the quantitative interpretation of geometry-dependent variability, especially in realistic vascular networks. However, the main reranking mechanisms identified in this work do not arise solely from source geometry: protocol-dependent changes already emerge under spatially uniform oxygen supply, through the coupling between oxygenation, radiosensitivity, and phenotypic selection. The heterogeneous-source analysis should therefore be interpreted as a controlled perturbation of this baseline mechanism, showing that spatial organization can further modulate both the expected gain and the reliability of protocols selected under simplified assumptions. Extensions to higher-dimensional and dynamically evolving vascular geometries may refine the quantitative magnitude of these effects, but are not expected to remove the underlying dependence of protocol ranking on oxygenation and phenotypic adaptation.

    Another simplifying assumption concerns tumor oxygen consumption, which is taken to be phenotype-independent. The present model does not explicitly resolve
    alternative metabolic pathways or phenotype-specific oxygen-consumption
    mechanisms. If, for example, hypoxia-adapted cells undergo metabolic
    reprogramming toward a more glycolytic phenotype and consequently consume less
    oxygen, their enrichment could reduce local oxygen depletion and promote
    reoxygenation. This additional feedback could modify the balance between
    hypoxic selection, proliferation, radiosensitivity, and treatment response,
    thereby shifting the quantitative predictions of the model. Other
    phenotype--metabolism relationships could lead to different feedbacks, and
    their impact would require an explicit phenotype-dependent description of
    oxygen uptake.

    A further limitation concerns the representation of phenotypic adaptation following reoxygenation. The present model does not explicitly account for \emph{hypoxic memory}, whereby cells may retain hypoxia-induced molecular and phenotypic programs for some time after reoxygenation \citep{Rocha2021,Sadhu2026}. The oxygen-dependent components of fitness and radiosensitivity at a given phenotype are determined by the current local oxygen level, and the local fitness optimum is correspondingly determined by this instantaneous oxygen environment (see Supporting Information S3). Although the effects of previous hypoxic exposure may persist indirectly through the evolving phenotypic composition of the tumor population, an explicit representation of hypoxic memory, as recently incorporated into a phenotype-structured PDE framework closely related to ours \citep{Sadhu2026}, could prolong hypoxia-associated behavior following reoxygenation and modify the quantitative balance between reoxygenation and hypoxia-driven selection that underlies treatment response and protocol ranking. This effect may be particularly relevant under the protracted, metronomic-like schedules favored under hypoxia. We regard this as a natural and promising model extension.

    Building on this framework, future developments may incorporate more realistic vascular architectures, dynamic oxygen delivery, and patient-specific information derived from functional imaging or multi-omics profiling. The model also lends itself to optimal control formulations in which fractionation schedules are optimized under explicit tumor control and normal-tissue objectives, and may be integrated with dose-painting strategies targeting spatially resolved regions, such as resistant niches that remain a subject of active debate~\citep{Qiu2017}.
    
    A complementary extension concerns uncertainty quantification. In the present study, variability is introduced through Monte Carlo realizations of oxygen-source geometry, while phenotypic dynamics, radiosensitivity parameters, and treatment response are otherwise deterministic for a given configuration. Further stochastic extensions could explicitly model intrinsic variability in phenotypic evolution, radiosensitivity, and treatment response, allowing the robustness of treatment comparisons to be assessed under broader biological uncertainty.

    Within these limitations, the proposed in silico approach offers a consistent quantitative setting for generating and testing hypotheses on how oxygenation, spatial heterogeneity, and phenotypic adaptation may affect radiotherapy schedule ranking. In particular, it provides a structured tool for exploring alternative fractionation strategies before considering more clinically constrained, patient-specific optimization settings.
\backmatter

\section*{Acknowledgements}
The authors are members of the Gruppo Nazionale per la Fisica Matematica (GNFM) of the Istituto Nazionale di Alta Matematica (INdAM).

\section*{Statements and Declarations}

\bmhead{Supporting information}

The supporting information is provided as a separate file.

\bmhead{Funding}
G.C. acknowledges financial support from the Ministry of Science and Innovation through the BCAM Severo Ochoa accreditation CEX2021-001142-S / MICIU / AEI / 10.13039/501100011033, as well as from the Basque Government through the IKUR Strategy (IKUR HPC\&IA 2025–2026) and from the European Union through NextGenerationEU/PRTR. This research was also supported by the Basque Government through the BERC 2022–2025 program. G.C. further acknowledges financial support from the European Research Council through the ERC Advanced Grant Nonlocal PDEs for Complex Particle Dynamics: Phase Transitions, Patterns and Synchronization (Grant Agreement No. 883363)..
M.D. was partially supported by MIUR CUP:
E11G18000350001.
The funders had no role in study design, data collection and analysis, decision to
publish, or preparation of the manuscript.

\bmhead{Competing interests}
The authors have no relevant financial or non-financial interests to disclose.

\bmhead{Ethics approval and consent to participate}
Not applicable.

\bmhead{Consent for publication}
Not applicable.

\bmhead{Data availability}
Not applicable.

\bmhead{Code availability}
The code used to perform the simulations and generate the results presented in this study is available at \url{https://staff.polito.it/marcello.delitala/code/submission_code.zip}.

\bmhead{Author contributions}
FA: Conceptualization, Methodology, Software, Formal analysis and investigation, Visualization, Writing -- original draft preparation, Writing -- review and editing. GC: Conceptualization, Methodology, Software, Formal analysis and investigation, Writing -- original draft preparation, Writing -- review and editing. MED: Conceptualization, Methodology, Formal analysis and investigation, Writing -- review and editing, Supervision.

\putbib[main_references_updated]
\end{bibunit}

%% SUPPORTING INFORMATION
\clearpage

\begin{bibunit}[sn-mathphys-ay]

\begin{center}
    {\Large\bfseries Supporting Information\par}

    \vspace{0.4cm}

    {\large
    Oxygenation and spatial heterogeneity shape radiotherapy protocol ranking through phenotypic adaptation
    \par}
   
    \vspace{0.4cm}

    {\normalsize
    Francesco Albanese$^{1,*}$, Giulia Chiari$^{2,3}$, Marcello Edoardo Delitala$^{1}$
    \par}

    \vspace{0.2cm}

    {\small
    $^{1}$ Department of Mathematical Sciences ``G. L. Lagrange'', Politecnico di Torino, Torino, Italy\\
    $^{2}$ Basque Center for Applied Mathematics (BCAM), Bilbao, Spain\\
    $^{3}$ Mathematical Institute, University of Oxford, Oxford, United Kingdom\\[0.15cm]
    $^{*}$ Corresponding author: francesco\_albanese@polito.it
    \par}
\end{center}

\vspace{0.6cm}

\setcounter{section}{0}
\setcounter{figure}{0}
\setcounter{table}{0}
\setcounter{equation}{0}

\renewcommand{\thesection}{S\arabic{section}}
\renewcommand{\thefigure}{S\arabic{figure}}
\renewcommand{\thetable}{S\arabic{table}}
\renewcommand{\theequation}{S\arabic{equation}}

This file contains supplementary material intended to support the results
presented in the main text, including additional derivations, figures,
and methodological details.

Throughout this Supporting Information, references to figures, equations,
tables, and sections without the prefix ``S'' refer to the main text.

    \section{Simulation details}
        \label{apx:simulation details}
        
            \subsection{Spatial domain and tumor burden}
            
            Simulations are performed on the one-dimensional spatial domain 
            $\Omega_s = [0,L_x]$ with $L_x = 2\,\mathrm{cm}$. 
            Recalling the volumetric interpretation of the cell density $n(t,x,u)$, the transverse dimensions are fixed to unit cross-sectional area ($1 \, \text{cm}^2$) when computing the total tumor burden, so that
            \begin{equation}
            \Gamma(t)
            =
            \int_{\Omega_s}\int_{\Omega_p}
            n(t,x,u)\,du\,dx
            \label{eq:total_burden}
            \end{equation}
            can be directly interpreted as the total number of tumor cells within the virtual tissue.

            \subsection{Detection threshold and simulation endpoints}
                All simulations start at $t=0$ and evolve until the tumor reaches a prescribed detection threshold. Following \citet{Erdi2012}, clinical detectability is typically associated with a minimal tumor spheroid of radius $r = 0.35$ cm. We convert this detectable volume into a total cell number by assuming an effective mean cellular packing equal to 70\% of the reference carrying capacity. This gives the detectability threshold
                
                \begin{equation}
                \label{eq:Gamma_RT}
                \Gamma_{\mathrm{RT}} = 1.3 \cdot 10^{8} \; \text{cells}.
                \end{equation}
                Radiotherapy is initiated when the total tumor burden exceeds this threshold:
                \begin{equation}
                t^{\mathrm{RT}}_0
                =
                \inf\left\{ t \;:\; \Gamma(t)\ge \Gamma_{\mathrm{RT}} \right\},
                \label{eq:rt_onset}
                \end{equation}
                and numerical simulations end at relapse time
                \begin{equation}
                t_{\mathrm{end}}
                =
                \inf\left\{ t>t^{\mathrm{RT}}_0 \;:\; \Gamma(t)\ge\Gamma_{\mathrm{RT}} \right\}.
                \label{eq:simulation_end}
                \end{equation}

            \subsection{Parameter estimates}   
                A large subset of the model parameters is informed by estimates available in the empirical literature. Accordingly, the adopted parameter configuration is intended to capture generic tumor behavior. When tumor-specific data are available, the model may be further calibrated to reflect the characteristics of a particular cancer type. The complete set of parameter values is reported in Table 1. 
                Whenever additional detail on parameter derivation or calibration is needed, the reader is referred to the relevant equations or to the sections where a full derivation is provided. 
        
        \subsection{Initial and boundary conditions}
            \label{par:ICBC}                
            At time \(t=0\), the tumor cell density is prescribed as a spatially localized distribution with phenotypes concentrated near the radiation-sensitive state:
            \begin{equation}
                n(0,x,u)=n_0(x,u):=A\,
                \exp\!\left(-\frac{(x-x_0)^2}{2\sigma_x^2}\right)\,
                \exp\!\left(-\frac{u^2}{2\sigma_u^2}\right),
                \qquad (x,u)\in\Omega_s\times\Omega_p,
            \end{equation}
            where \(x_0=L_x/2\) denotes the initial tumor location, while \(\sigma_x>0\) and \(\sigma_u>0\) control the spatial and phenotypic spread, respectively. The amplitude \(A=10^5\,\text{cell}/\text{cm}^3\) is chosen so that the initial tumor density lies well below the prescribed carrying-capacity scale \(K\) throughout the spatial domain. This ensures that the system initially evolves in a low-density regime, before density-dependent crowding becomes relevant. The calibration of the crowding coefficient \(\kappa\) based on \(K\) is described in Section~\ref{apx:rho_star}.
            
            The initial oxygen concentration is prescribed through a consistent initial profile. Since oxygen dynamics evolve on a timescale that is several orders of magnitude faster than tumor cell proliferation, the specific choice of the initial condition has a negligible impact on the subsequent dynamics.
            
            No-flux boundary conditions are imposed both in space and in phenotype for the tumor cell density, and in space for the oxygen field. Biologically, these conditions represent physical and functional barriers that prevent cell escape from the considered tissue portion and exclude phenotypic flux across the extreme resistance states, while assuming no external oxygen exchange across the tissue boundaries. 
                                    
        \subsection{Numerical methods}
            \label{subsubsec:numerical_methods}
            
            The spatio--phenotypic system for the tumor cell density $n(t,x,u)$ and the
            oxygen field $O(t,x)$, defined in
            Eqs.\,(1)-(9), is solved numerically
            by finite differences on uniform grids in space and phenotype.
            Diffusion operators are approximated by second-order centered finite
            differences.
            Time integration is performed using a first-order explicit (forward Euler) scheme combined with operator splitting of reaction, phenotypic diffusion, and spatial diffusion terms.
            The oxygen equation is advanced by an explicit reaction-diffusion update,
            with internal substepping when required by stability constraints.
            Radiotherapy is implemented in an event-driven fashion: the time step is
            adaptively adjusted to land exactly on scheduled dosing times, at which the
            active cell population is instantaneously updated according to the modified
            linear-quadratic survival model proposed in Eq.\,(5).

        \subsection{Computational details} 
            \label{sec:S_computational_details}
            All simulations were implemented in \textsc{MATLAB} R2024b. The spatial and phenotypic domains were discretized using uniform grids with mesh sizes \(\Delta x=2 \, \mathrm{cm}/100\) and \(\Delta u=1/100\), respectively. One ghost point was included at each boundary to impose no-flux boundary conditions. The oxygen field was discretized on the same spatial grid. Integral quantities were computed using native trapezoidal-rule routines.
        
            Given the large number of independent model evaluations required by the dose--interval scans and by the Monte Carlo exploration of heterogeneous oxygen-source configurations, simulations were parallelized at the level of independent runs. Each dose--interval map was obtained from \(50\times 50=2500\) independent simulations, corresponding to the explored combinations of single-fraction dose and inter-fraction interval. In the heterogeneous-source analysis, each set of \(2500\) protocols was evaluated over \(100\) independently generated oxygen-source configurations, yielding \(2500\times 100\) simulations for each oxygenation regime considered.
        
            Large batches of simulations were executed on a high-performance computing cluster using the \textsc{SLURM} workload manager. This parallelization did not alter the numerical scheme described above: each job corresponded to an independent realization of a prescribed protocol and oxygen-source configuration. The computational cost varied across regimes, mainly because oxygenation level, spatial heterogeneity, and treatment response affect the simulated time-to-progression and therefore the number of time steps required before relapse. A typical single simulation required approximately \(2\) minutes in normoxic or weakly heterogeneous settings, and up to approximately \(10\) minutes in hypoxic and/or highly heterogeneous configurations. Thus, the exploration of dose--interval protocols and oxygen-source geometries was computationally intensive but naturally suited to parallelization, since individual simulations were independent.
    \section{Indicator definitions} 
        \label{apx:indicator definitions}
        
        \subsection{Biologically Effective Dose (BED)}
            \label{apx:BED}
            The \emph{biologically effective dose} (BED) associated with a fractionation protocol 
            $\mathcal{P}=(d,\Delta,N_f)$ is defined, according to the linear--quadratic formulation and its application to fractionated radiotherapy \citep{Fowler1989}, as
            \begin{equation}
            \label{eq:BED}
            \mathrm{BED}(\mathcal{P}) 
            = 
            N_f\, d \left( 1 + \frac{d}{(\alpha/\beta)_H} \right),
            \end{equation}
            where $N_f$ is the number of fractions and $d$ the dose per fraction (Gy). 
            The $(\alpha/\beta)_H$ ratio characterizes the fractionation sensitivity of the surrounding healthy tissue, with larger values associated with early-responding tissues and smaller values with late-responding tissues \citep{Freedman2013}. 
            The BED provides a scalar measure of the expected biological effect of a fractionation schedule and is here used to quantify normal-tissue toxicity, enabling comparison across protocols under a common constraint.
                        
        \subsection{Mean phenotype}
            \label{apx:u_mean}
            To quantify tumor adaptation to therapy and to microenvironmental stress induced by hypoxia, the \emph{mean phenotype} is introduced:
            \begin{equation}
                \bar{u}(t) =
                \frac{\displaystyle \int_{\Omega_p} \int_{\Omega_s} u\, n(t,x,u)\,\mathrm{d}x\,\mathrm{d}u}
                     {\displaystyle \int_{\Omega_s} \rho(t,x)\,\mathrm{d}x}.
            \end{equation}
            This quantity represents the average resistance level of the tumor population at time~$t$, accounting for both spatial structure and phenotypic heterogeneity.
            This and similar averaged phenotypic indicators have been widely used in phenotypically structured models to characterize the macroscopic evolutionary dynamics of heterogeneous populations; see, for instance, \citet{Lorenzi2025} and references therein.

        \subsection{Time-to-progression (TTP)}
            \label{par:TTP}
            The performance of a treatment protocol is quantified through the \emph{time-to-progression} (TTP), which serves as the primary efficacy indicator in our simulations. TTP measures the time interval between treatment initiation and tumor relapse:
            \begin{equation}
            \mathrm{TTP}
            :=
            t_{\mathrm{end}} - t^{\mathrm{RT}}_0.
            \end{equation}
            Such metrics have been used in previous studies to assess and compare treatment strategies (see, e.g., \citealp{HenaresMolina2017,Bruningk2021}).

    \section{Existence and uniqueness of the fittest phenotype}
        \label{apx:u*}
        For each spatial location $x \in \Omega_s$ and time $t \in [0,t_{\text{end}}]$, the local oxygen
        concentration $O(t,x)$ induces a fitness landscape over the phenotypic trait $u \in \Omega_p = [0,1]$
        through the reaction term $R(t,x,u)$ in Eq.\,(2). Fixing $(t,x)$, we can regard the reaction term as a
        function of $u$ only and write
        
        \begin{equation}
            R(u; t,x)
            = r_0 + \gamma_O\,M\big(O(t,x)\big)\,\big(1-u^2\big)
             - \gamma_H\big(1-M\big(O(t,x)\big)\big)\,\big(1-u\big)^2
             - \kappa\,\rho(t,x),
            \label{eq:appendixA_R}
        \end{equation}
        where $M(\cdot)$ is the Hill-type function introduced in Eq.\,(3) and $\rho(t,x)$ is the local total cell density defined in Eq.\,(4).
        Since the crowding term $-\kappa\,\rho(t,x)$ does not depend on $u$, the phenotype that maximizes
        $R(u; t,x)$ coincides with the maximizer of
        
        \begin{equation}
            \widetilde R(u; t,x)
            = r_0 + \gamma_O\,M\big(O(t,x)\big)\,\big(1-u^2\big)
              - \gamma_H\big(1-M\big(O(t,x)\big)\big)\,\big(1-u\big)^2.
            \label{eq:appendixA_Rtilde}
        \end{equation}
        For notational brevity, in what follows we denote $M = M\big(O(t,x)\big)$ and omit the explicit
        dependence on $(t,x)$.
        The first derivative of $\widetilde R$ with respect to $u$ is
        \begin{equation}
            \frac{d}{du}\widetilde R(u)
            = -2\,\gamma_O\,M\,u
              + 2\,\gamma_H\,(1-M)\,(1-u).
        \end{equation}
        The critical point $u^\ast$ is obtained by solving $\frac{d}{du}\widetilde R(u)=0$, which yields
        \begin{equation}
        u^\ast
        = \frac{\gamma_H (1-M)}{\gamma_O M + \gamma_H (1-M)}.
        \label{eq:appendixA_ustar}
        \end{equation}
        Using the assumptions $\gamma_O>0$, $\gamma_H>0$ and $M \in [0,1]$, it follows that
        $0 \le \gamma_H(1-M) \le \gamma_O M + \gamma_H(1-M)$, so that $u^\ast \in [0,1]$.
        Therefore, $u^\ast$ is an admissible phenotype in $\Omega_p$.
        To show that $u^\ast$ is the unique global maximizer of $R(u;t,x)$ with respect to $u$, we compute
        the second derivative:
        \begin{equation}
        \frac{d^2}{du^2}\widetilde R(u)
        = -2\big(\gamma_O M + \gamma_H(1-M)\big) < 0
        \quad \text{for all } u \in [0,1],
        \end{equation}
        which confirms that $\widetilde R(u)$ (and hence $R(u;t,x)$) is strictly concave in $u$. As a
        consequence, $u^\ast$ defined in~\eqref{eq:appendixA_ustar} is the unique fittest phenotype at
        position $x$ and time $t$.
        The dependence of $u^\ast$ on the oxygen concentration $O(t,x)$ is encoded through the function
        $M\big(O(t,x)\big)$, which is strictly increasing in $O$. In particular,
        \begin{equation}
        O \to 0 \;\Rightarrow\; M(O) \to 0 \;\Rightarrow\; u^\ast \to 1,
        \qquad
        O \to +\infty \;\Rightarrow\; M(O) \to 1 \;\Rightarrow\; u^\ast \to 0.
        \end{equation}
        Therefore, in highly hypoxic regions the selected phenotype tends toward $u^\ast \approx 1$, i.e.
        toward highly resistant traits, whereas in well-oxygenated regions it tends toward $u^\ast \approx 0$,
        corresponding to more proliferative and less resistant phenotypes. This behavior is consistent with
        the modeling interpretation of $u$ adopted in the main text and with the eco-evolutionary role of
        hypoxia in selecting resistant tumor cell clones.

    \section{Calibration of the crowding coefficient}
        \label{apx:rho_star}
        
        For fixed $(t,x,u)$, the net proliferation rate can be written as
        \begin{equation}
        R(t,x,u)
        =
        F\bigl(O(t,x),u\bigr)-\kappa\rho(t,x),
        \end{equation}
        where
        \begin{equation}
        F(O,u)
        =
        r_0
        +
        \gamma_O M(O)(1-u^2)
        -
        \gamma_H\bigl(1-M(O)\bigr)(1-u)^2 .
        \end{equation}
        Since \(M(O)\in[0,1)\) and \(u\in[0,1]\), the intrinsic growth term satisfies
        \begin{equation}
        \sup_{O,u} F(O,u)=r_0+\gamma_O .
        \end{equation}
        We calibrate \(\kappa\) so that, when the local density reaches \(K\), even the most favorable growth conditions do not yield positive net proliferation:
        \begin{equation}
        r_0+\gamma_O-\kappa K=0.
        \end{equation}
        Accordingly,
        \begin{equation}
        \kappa=\frac{r_0+\gamma_O}{K}.
        \label{eq:kappa_calibration}
        \end{equation}
        In this sense, \(K\) is interpreted as an upper-bound carrying capacity.
        Throughout the model, the cell density \(n(t,x,u)\) is interpreted as
        volumetric (cells/cm\(^3\)). We set \(K=10^9\) cells/cm\(^3\), corresponding
        to the typical number of cells in a \(1\,\mathrm{cm}^3\) tumor~\citep{DelMonte2009}.
    
    \section{An integrated perspective on cellular radiosensitivity}
        \label{apx:radio}

        In the LQ framework, the parameters $\alpha$ and $\beta$ quantify cell death from single- and multiple-track events \citep{McMahon2018}. Reported values show substantial variability across tumor types \citep{vanLeeuwen2018}, which is typically larger than the ranges assumed in modeling studies.
        The ratio $\alpha/\beta$ represents the dose at which linear and quadratic contributions are equal and is therefore relevant for fractionation sensitivity.
        
        Oxygen effects are commonly incorporated through the Oxygen Enhancement Ratio (OER), defined as the factor by which radiation dose must be increased under hypoxia to achieve the same biological effect as in normoxia. This leads to the rescaling
        \begin{equation}
        \alpha_{\mathrm{norm}} = \mathrm{OER}\,\alpha_{\mathrm{hyp}}, 
        \quad
        \beta_{\mathrm{norm}} = \mathrm{OER}^2\,\beta_{\mathrm{hyp}},
        \end{equation}
        implying $\alpha_{\mathrm{hyp}}/\beta_{\mathrm{hyp}} = \mathrm{OER}\,\alpha_{\mathrm{norm}}/\beta_{\mathrm{norm}}$. Typical values of OER range up to approximately $2.5$--$3$ depending on oxygenation level.
        However, experimental evidence is not fully consistent with this prediction. While in vitro studies support the OER-based scaling \citep{Stea2012}, clinical observations often report lower $\alpha/\beta$ ratios in hypoxic tumors, which are frequently treated with hypofractionated schedules \citep{Wang2003,Mohamed2023,Ritter2009}. In addition, hypoxia affects DNA repair and promotes resistant, stem-like phenotypes.
        To account for these effects, we define $\alpha$ and $\beta$ as functions of oxygen concentration and phenotype in Eq.\,(6). Starting from baseline values $\tilde{\alpha}$ and $\tilde{\beta}$, we introduce phenotype-dependent amplitudes $\Delta_\alpha$ and $\Delta_\beta$, combined with an oxygen-dependent rescaling consistent with the OER formulation. Parameters are chosen so that, for $\text{OER}_{\max}=3$, the resulting values of $\alpha$ and $\beta$ span the ranges reported in the literature \citep{vanLeeuwen2018}.
        
\begin{figure*}[h!]
            \centering
            \captionsetup{font=small}
            \includegraphics[width=\textwidth]{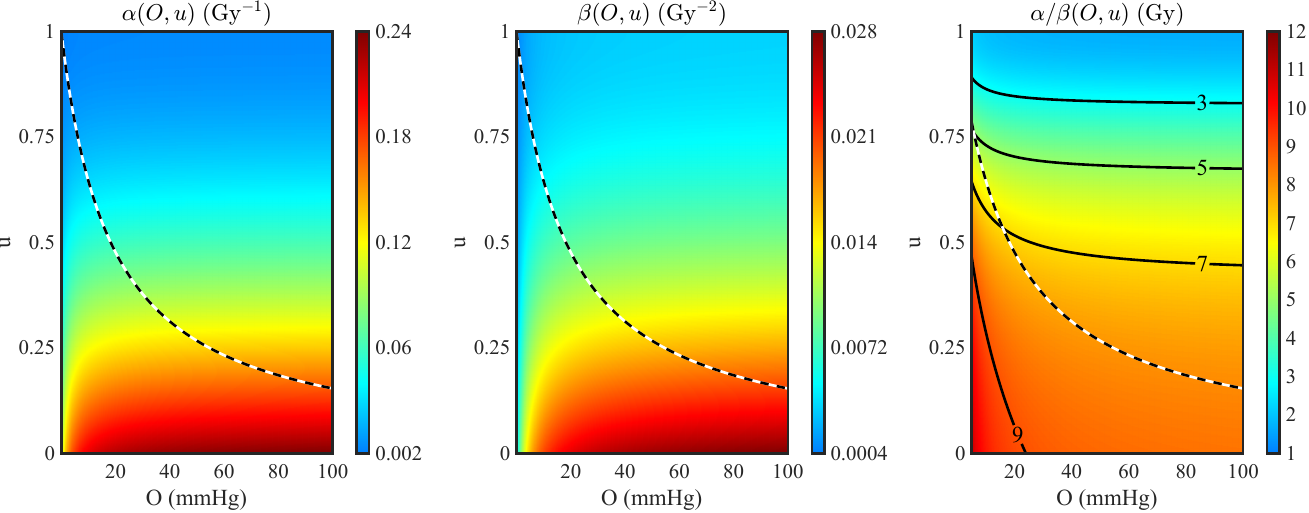}
            \caption{
            Influence of oxygen concentration and phenotypic state on the radiosensitivity
            parameters of the modified LQ model defined in Eq.\,(6). \textbf{Left:} values of
            \(\alpha(O,u)\). \textbf{Center:} values of \(\beta(O,u)\). \textbf{Right:} corresponding
            \(\alpha/\beta\) ratio, with contour lines highlighting constant-ratio levels.
            The dotted curve denotes the fittest phenotype
            \(u=u^*\) (see Section~\ref{apx:u*}) as a function of oxygen concentration.
            Although decreasing oxygen increases \(\alpha/\beta\) at fixed phenotype, the
            ratio decreases along the \(u^*\) trajectory because hypoxia simultaneously
            selects for more resistant phenotypes.
            }
            \label{Fig.:OER}
        \end{figure*}

        In this setting, hypoxia directly reduces radiosensitivity through the standard OER mechanism, while hypoxia-driven selection shifts the population toward more resistant phenotypes.
        The combined effect of oxygenation and phenotypic composition modifies both radiosensitivity and the $\alpha/\beta$ ratio. At fixed phenotype, decreasing oxygen increases $\alpha/\beta$, as predicted by the classical formulation. Along the phenotype selected by the system dynamics (see Supporting Information \ref{apx:u*}), the ratio instead decreases toward hypoxia due to the concurrent shift toward resistant traits (Fig.~\ref{Fig.:OER}).
        This formulation provides a unified description that captures both controlled conditions, where oxygen and phenotype can be varied independently (\textit{in vitro}), and evolving populations subject to selection (\textit{in vivo}).
    
    \section{Oxygen dynamics quantification: baseline consumption and tumor uptake}
        \label{apx:oxypar}
        This section details the derivation and calibration of the oxygen decay and consumption parameters used in Eq.~(9). The goal is to relate the effective coefficients $\lambda_O$ and $\zeta_O$ to physiological quantities reported in the literature.
        To parameterize the baseline oxygen decay term $-\lambda_O O$, we relate it to the physiological oxygen consumption of healthy tissue. Assuming homeostatic conditions, the density of healthy cells is fixed at the carrying capacity 
        $K_H = 5 \cdot 10^{8}\,\mathrm{cells/cm^3}$.
        We denote by $\tilde{\zeta}_H$ the per-cell oxygen consumption rate,
        $\tilde{\zeta}_H = 5 \cdot 10^{-18}\,\mathrm{mol\,cell^{-1}\,s^{-1}} = 4.32 \cdot 10^{-13}\,\mathrm{mol\,cell^{-1}\,day^{-1}}$,
        within the range reported in~\citet{Wagner2011}. Using the molar volume of oxygen 
        $\eta_O = 2.24 \cdot 10^{4}\,\mathrm{ml\,O_2/mol}$
        and a reference normoxic level 
        $O_{\mathrm{ref}} = 2.16 \cdot 10^{-3} \,\mathrm{ml\,O_2/cm^3}$,
        we define the volumetric consumption coefficient as 
        $\zeta_H := \tilde{\zeta}_H \eta_O / O_{\mathrm{ref}}$, 
        which yields
        \begin{equation}
        \label{eq:lambda_O}
        \lambda_O = \zeta_H K_H.
        \end{equation}
        This corresponds to a linearization of volumetric oxygen consumption around the reference normoxic state.
        The above quantities are expressed in the units adopted in the referenced study. The resulting parameter is defined consistently with the model formulation, and its final value is reported in Table~1.
        
        Oxygen consumption by tumor cells is modeled through the nonlinear term $-\zeta_O \rho O$, where $\rho$ denotes the tumor cell density. The consumption rate is assumed phenotype-independent.
        Neglecting diffusion, one may write $\partial_t O = V_O(x) - \zeta_H K_H O - \zeta_T \rho O$, where $\zeta_T$ is the tumor per-cell consumption coefficient. Since the term $-\lambda_O O$ already accounts for healthy tissue consumption, the tumor contribution is interpreted as an excess uptake due to the replacement of healthy cells. Assuming a higher compressibility of tumor cells, allowing for up to a one-to-two replacement ratio,
        %ne healthy cell is replaced by two tumor cells \misscit{},
        we obtain the effective formulation
        \begin{equation}
        \partial_t O = V_O(x) - \lambda_O O - \zeta_O \rho O,
        \end{equation}
        with $\zeta_O = \zeta_T - \zeta_H/2$.
        Accordingly, we adopt a per-cell tumor consumption rate 
        $\tilde{\zeta}_T = 6.7 \cdot 10^{-17}\,\mathrm{mol\,cell^{-1}\,s^{-1}} = 5.79 \cdot 10^{-12}\,\mathrm{mol\,cell^{-1}\,day^{-1}}$,
        within the range reported in~\cite{Wagner2011}, yielding the coefficient $\zeta_O$ listed in Table 1.
        
    \section{Quantification of the therapeutic efficiency frontier}
        \label{apx:barycenter}
        
        We consider the set of fractionated radiotherapy protocols used to construct
        the dose--interval maps. Each protocol consists of identical dose fractions
        delivered at a constant inter-fraction interval, with the number of fractions
        determined by the imposed BED constraint.
        For each \(i\in\mathcal{I}\), where \(\mathcal{I}\) denotes the index set of
        all protocols explored, we denote by
        \(\mathcal{P}_i\) the \(i\)-th simulated protocol. The corresponding
        dose--interval pair is \((d_i,\Delta_i)\), and \(\mathrm{TTP}_i\) denotes the
        time-to-progression obtained from the simulation. We define
        \begin{equation}
            \mathrm{TTP}_{\max} := \max_{i\in\mathcal{I}} \mathrm{TTP}_i
        \end{equation}
        as the maximum time-to-progression attained in that protocol set.
        The \emph{therapeutic efficiency frontier} is then defined as the set of
        protocols
        \begin{equation}
        \mathcal{F}
        := \left\{\, \mathcal{P}_i \;:\; i\in\mathcal{I},
        \ \mathrm{TTP}_i \geq 0.8\cdot\mathrm{TTP}_{\max} \,\right\}.
        \label{eq:appendixB_frontier}
        \end{equation}
        Thus, \(\mathcal{F}\) identifies the simulated protocols achieving at least
        \(80\%\) of the maximum predicted TTP. It can therefore be interpreted as the
        set of near-optimal protocols with respect to this performance metric.
        
        To provide a compact summary of the typical location of the therapeutic
        efficiency frontier in the \((d,\Delta)\) plane, we introduce its
        TTP-\emph{weighted barycenter}. Specifically, the barycenter
        \((\bar d,\bar\Delta)\) is defined as
        \begin{equation}
        \bar d
        := \frac{\displaystyle \sum_{\mathcal{P}_i\in\mathcal{F}} d_i\,\mathrm{TTP}_i}
                {\displaystyle \sum_{\mathcal{P}_i\in\mathcal{F}} \mathrm{TTP}_i},
        \qquad
        \bar\Delta
        := \frac{\displaystyle \sum_{\mathcal{P}_i\in\mathcal{F}} \Delta_i\,\mathrm{TTP}_i}
                {\displaystyle \sum_{\mathcal{P}_i\in\mathcal{F}} \mathrm{TTP}_i}.
        \label{eq:appendixB_barycenter}
        \end{equation}
        By construction, protocols yielding larger TTP values contribute more strongly
        to the location of the barycenter. The barycenter is not intended to identify
        an optimal treatment schedule. Rather, it provides a low-dimensional descriptor
        of the region in the dose--interval space associated with highly effective
        regimens. This representation facilitates comparisons across different
        oxygenation conditions and toxicity constraints, and is therefore used
        throughout the manuscript to track qualitative shifts in the therapeutic
        efficiency frontier.
    
    \section{Construction of heterogeneous oxygen sources}
        \label{apx:heterogeneous_sources}

        \subsection{Monte Carlo construction of oxygen sources}
            To investigate the effect of spatial heterogeneity in oxygen delivery, we construct an ensemble of oxygen source configurations with identical total input and varying spatial organization. The ensemble is generated through a Monte Carlo procedure, so that each realization corresponds to a different spatial redistribution of the same total oxygen supply.
            In the spatially uniform case, the oxygen source is defined as
            \begin{equation}
            V_O(x) = I_0 \,\lambda_O,
            \qquad x \in \Omega_s = [0,L_x],
            \end{equation}
            which yields a constant oxygen supply across the domain.
            In the heterogeneous setting, for each Monte Carlo realization \(m=1,\dots,M\), we construct a raw oxygen source profile as a superposition of \(n_V\) localized Gaussian-shaped contributions:
            \begin{equation}
            V_{\mathrm{raw}}^{(m)}(x)
            =
            \sum_{i=1}^{n_V} w_i^{(m)}
            \exp\!\left(
            -\frac{\left(x-c_i^{(m)}\right)^2}
            {2\left(\sigma_i^{(m)}\right)^2}
            \right),
            \qquad x\in\Omega_s.
            \end{equation}
            The parameter \(n_V\) controls the number of localized oxygen-releasing regions and therefore the spatial granularity of oxygen delivery. It should not be interpreted as the literal number of physical vessels, but rather as a modeling parameter determining the complexity of the oxygen source.
            For each realization, the centers \(c_i^{(m)}\) are sampled independently and uniformly over the spatial domain,
            \begin{equation}
            c_i^{(m)} \sim \mathcal{U}[0,L_x],
            \end{equation}
            while the widths \(\sigma_i^{(m)}\) are sampled independently from a uniform distribution over a prescribed range. These widths do not have a direct biological interpretation and are introduced as a modeling choice to generate a broad variety of spatial configurations.
            The relative amplitudes \(w_i^{(m)}\) are sampled independently from a uniform
            distribution.
            Since the raw profile \(V_{\mathrm{raw}}^{(m)}\) generally has a realization-dependent integral, we apply a realization-dependent scaling factor to enforce conservation of the total oxygen supply:
            \begin{equation}
            V_O^{(m)}(x)
            =
            \eta^{(m)} \, V_{\mathrm{raw}}^{(m)}(x),
            \end{equation}
            where
            \begin{equation}
            \eta^{(m)}
            =
            \frac{I_0\,L_x\,\lambda_O}
            {\displaystyle\int_{\Omega_s} V_{\mathrm{raw}}^{(m)}(x)\,dx}.
            \end{equation}
            With this normalization, all realizations satisfy
            \begin{equation}
            \int_{\Omega_s} V_O^{(m)}(x)\,dx = I_0\,L_x\,\lambda_O,
            \end{equation}
            so that they share the same total oxygen supply and differ only in the spatial geometry of delivery.
            This construction allows us to isolate the effect of spatial heterogeneity on treatment outcomes by comparing different geometric configurations under identical global oxygen supply.
        
        \subsection{Quantification of spatial heterogeneity}
            To provide a compact descriptor of spatial heterogeneity, we associate to each
            oxygen-source realization \(m\) a normalized Gini coefficient
            \(\mathcal{G}^{(m)}\), computed from the corresponding discretized source
            profile \(V_O^{(m)}(x)\). Let
            \begin{equation}
            v^{(m)}=\left(v^{(m)}_1,\dots,v^{(m)}_n\right)
            \end{equation}
            denote the values of \(V_O^{(m)}(x)\) on the spatial grid, normalized so that
            \begin{equation}
            \sum_{k=1}^n v^{(m)}_k = 1.
            \end{equation}
            Here, \(v^{(m)}_k\) denotes the value at the \(k\)-th spatial grid point. Let
            \begin{equation}
            v^{(m)}_{(1)}\leq v^{(m)}_{(2)}\leq \dots \leq v^{(m)}_{(n)}
            \end{equation}
            denote the same values sorted in non-decreasing order. The normalized Gini
            coefficient of realization \(m\) is then defined as
            \begin{equation}
            \mathcal{G}^{(m)}
            =
            \frac{1}{n-1}
            \sum_{k=1}^n (2k - n - 1)\, v^{(m)}_{(k)}.
            \end{equation}
            The normalization by \(n-1\) ensures that
            \(\mathcal{G}^{(m)}\in[0,1]\) on the finite spatial grid. By construction,
            \(\mathcal{G}^{(m)}=0\) corresponds to a perfectly uniform oxygen source,
            whereas \(\mathcal{G}^{(m)}=1\) corresponds to the limiting case in which all
            oxygen supply is concentrated at a single grid point. Therefore, the Gini coefficient primarily captures the degree of \emph{polarization} of
            the oxygen source, i.e., how strongly oxygen delivery is concentrated in
            localized regions. In this work, it is adopted as a first, compact descriptor
            of spatial heterogeneity. Other indicators could in principle be used to
            characterize different aspects of the spatial organization of the oxygen
            distribution.

    \section{Schedule-domain robustness analysis}
        \label{apx:clinically_constrained}
    
        The work presented in \citet{HenaresMolina2017} represents a key contribution
        to the study of optimal radiotherapy protocols while also raising important
        open questions. Our work builds on these aspects by further investigating how
        the evolutionary and environmental features of the tumor influence the selection
        of optimal treatment strategies. We have shown that oxygenation and tumor
        heterogeneity strongly affect the benefit of alternative protocols over the
        standard-of-care.
        
        We further examine robustness with respect to the explored schedule domain by
        repeating the dose--interval scans under progressively restricted admissible
        inter-fraction interval ranges, while keeping the BED constraint unchanged. We
        test whether the main oxygenation-dependent patterns reported in the
        manuscript---including the emergence of larger gains away from SoC in
        intermediate hypoxia---persist when the maximum allowable spacing between
        fractions is reduced. Specifically, we considered
        \begin{equation}
            \Delta_{\max}\in\{7,14,30,60,100\}\ \mathrm{days}
        \end{equation}
        where \(\Delta_{\max}=100\) corresponds to the extended range used in the
        main-text dose--interval maps.
        Here, \(I_0\) denotes the reference tumor-free oxygenation level used to set the
        total oxygen supply. For each pair \((I_0,\Delta_{\max})\), let
        \(\mathcal{I}(I_0;\Delta_{\max})\) denote the index set of all simulated
        protocols in the restricted scan, with \(\mathcal{P}_i\) denoting the \(i\)-th
        simulated protocol. The restricted scan contains only protocols satisfying the
        same BED constraint used in the main text and the additional condition
        \(\Delta_i\leq\Delta_{\max}\).
        For each restricted scan, we first define the maximum TTP attained over the
        admissible protocol set as
        \begin{equation}
        \mathrm{TTP}_{\max}(I_0;\Delta_{\max})
        :=
        \max_{i\in\mathcal{I}(I_0;\Delta_{\max})}
        \mathrm{TTP}_i(I_0;\Delta_{\max})
        \end{equation}
        The corresponding therapeutic efficiency frontier is then defined as
        \begin{equation}
        \mathcal{F}(I_0;\Delta_{\max})
        :=
        \left\{
        \mathcal{P}_i :
        i\in\mathcal{I}(I_0;\Delta_{\max}),
        \ 
        \mathrm{TTP}_i(I_0;\Delta_{\max})
        \geq
        0.8 \cdot \mathrm{TTP}_{\max}(I_0;\Delta_{\max})
        \right\}
        \end{equation}
        This definition follows Section~\ref{apx:barycenter}, with the admissible set
        now restricted by \(\Delta_i\leq\Delta_{\max}\).
        
        For each protocol in this frontier, the gain over standard-of-care is
        computed by comparing its TTP with the SoC TTP under the same oxygenation level.
        We therefore define the median gain as
        \begin{equation}
        \widetilde{G}(I_0;\Delta_{\max})
        :=
        \operatorname{median}_{\mathcal{P}_i\in\mathcal{F}(I_0;\Delta_{\max})}
        \left(
        \mathrm{TTP}_i(I_0;\Delta_{\max})
        -
        \mathrm{TTP}_{\mathrm{SoC}}(I_0)
        \right)
        \end{equation}
        where \(\mathrm{TTP}_{\mathrm{SoC}}(I_0)\) denotes the TTP obtained with the
        standard-of-care schedule at the same oxygenation level. Since the SoC schedule
        has \(\Delta_{\mathrm{SoC}}=1\) day, it remains admissible for all restricted
        domains considered here. Hence, the SoC baseline depends on \(I_0\), but not on
        \(\Delta_{\max}\).
        We also compute the median TTP over the same frontier,
        \begin{equation}
        \widetilde{\mathrm{TTP}}(I_0;\Delta_{\max})
        :=
        \operatorname{median}_{\mathcal{P}_i\in\mathcal{F}(I_0;\Delta_{\max})}
        \mathrm{TTP}_i(I_0;\Delta_{\max})
        \end{equation}
        which provides the typical performance of the near-optimal protocols.
        In addition, we compute the TTP-weighted frontier barycenter
        \((\bar d,\bar\Delta)\) as in Section~\ref{apx:barycenter}. Finally, to compare
        relative improvements across oxygenation levels with different SoC baselines,
        we also report the median gain normalized by
        \(\mathrm{TTP}_{\mathrm{SoC}}(I_0)\). 
        
        Fig.~S2 summarizes these complementary aspects of the restricted-frontier
        analysis.
        
        The top-left panel reports the median gain over SoC across protocols in
        \(\mathcal{F}(I_0;\Delta_{\max})\). The gain depends non-monotonically on
        oxygenation, with the largest improvements in the intermediate hypoxic range.
        Restricting \(\Delta_{\max}\) reduces the attainable median gains, especially
        for short admissible intervals, while preserving the same qualitative
        oxygenation-dependent trend.
        
        The top-right panel reports the TTP-weighted barycenters
        \((\bar d,\bar\Delta)\) of the corresponding frontiers. These barycenters
        identify the typical location of the near-optimal region in the dose--interval
        plane. At intermediate oxygenation, the frontier shifts toward lower
        per-fraction doses and longer inter-fraction intervals. As \(\Delta_{\max}\)
        decreases, this shift is progressively constrained by the imposed upper bound
        on the admissible spacing between fractions. Because the frontier is recomputed
        within each restricted schedule domain, this constraint affects both components
        of the barycenter: in addition to reducing \(\bar\Delta\), it also tends to move
        the near-optimal region toward lower values of \(\bar d\), particularly in the intermediate hypoxic range.
        
        The bottom-left panel reports the median TTP of protocols in
        \(\mathcal{F}(I_0;\Delta_{\max})\), together with the corresponding SoC baseline. The median TTP of near-optimal schedules peaks at intermediate oxygenation, whereas near normoxia it approaches the SoC baseline, consistently with the flatter dose--interval landscape observed in the main text.
        
        The bottom-right panel reports the same median gain shown in the top-left panel
        after normalization by the corresponding SoC baseline. This representation
        shows that the proportional benefit of schedule adaptation is maximal at
        intermediate hypoxia and becomes modest near normoxia and under severe hypoxia.
        
        Overall, these results show that the qualitative conclusions reported in the
        main text are robust to the choice of the explored inter-fraction interval
        range. The same oxygenation-dependent patterns persist under progressively
        restricted \(\Delta_{\max}\), although the attainable magnitude of the gain and
        the accessible region of the dose--interval plane are reduced when the maximum
        allowed spacing between fractions is shortened.

    \section{Hill-coefficient sensitivity analysis}
        \label{sec:S10_hill_sensitivity}
        
        {
        
        The Hill-type function introduced in Eq.~(3) modulates the
        oxygen-dependent proliferative contribution and can be written more generally as
        \begin{equation}
            M(O;n_H)
            =
            \frac{O^{n_H}}
            {O^{n_H}+\alpha_O^{n_H}},
            \label{eq:S_hill_general}
        \end{equation}
        where \(n_H>0\) controls the steepness of the transition between low- and
        high-oxygen conditions. In the main analysis, we set \(n_H=4\) as a modeling choice, providing a sufficiently sharp transition and an effectively saturated response within the normoxic range.
        
        Changing \(n_H\) modifies the sharpness of the oxygen response without altering
        either its limiting values or its half-saturation point, since
        \begin{equation}
            M(\alpha_O;n_H)=\frac{1}{2}
            \label{eq:S_hill_half}
        \end{equation}
        for every \(n_H\). The Hill coefficient therefore determines how rapidly the
        response approaches its low- and high-oxygen plateaus while leaving the
        characteristic transition oxygen level unchanged.
        
        This distinction is particularly relevant for small Hill coefficients. At the
        reference normoxic oxygen level \(O_M=72.2\)~mmHg, the response is approximately
        \begin{equation}
            M(O_M;1)=0.826,
            \qquad
            M(O_M;2)=0.958,
            \qquad
            M(O_M;3)=0.991,
            \label{eq:S_hill_saturation}
        \end{equation}
        whereas \(M(O_M;n_H)>0.998\) for \(n_H\geq4\). Thus, \(n_H=1\), and to a
        lesser extent \(n_H=2\), produces a substantially more gradual transition and
        does not reproduce the same effective saturation of the oxygen-dependent
        proliferative contribution under normoxic conditions. Conversely,
        \(n_H\geq3\) yields an oxygen response that is already effectively saturated
        within the normoxic range while still spanning different transition steepnesses
        around \(O=\alpha_O\).
        
        To determine whether the treatment-selection results depend on this modeling
        choice, we repeated the complete dose--interval analysis underlying the
        main-text treatment landscapes for
        \begin{equation}
            n_H\in\{1,2,3,4,5,6\},
            \label{eq:S_hill_values}
        \end{equation}
        under both the normoxic condition \(I_0=O_M\) and the hypoxic condition
        \(I_0=O_m\). The analysis was performed in the spatially uniform oxygen-source setting used for main-text Fig.~1; spatial heterogeneity in the oxygen source was not considered here. All remaining model parameters, simulation settings, the BED constraint, and the dose--interval protocol domain were kept unchanged.
        
        Following the definitions introduced in Section~S7, for each Hill coefficient
        and oxygenation condition we independently define the corresponding \emph{therapeutic
        efficiency frontier} as
        \begin{equation}
            \mathcal{F}_{n_H}(I_0)
            =
            \left\{
                \mathcal{P}:
                \mathrm{TTP}_{n_H}(\mathcal{P};I_0)
                \geq
                0.8\cdot
                \mathrm{TTP}^{\max}_{n_H}(I_0)
            \right\},
            \label{eq:S_hill_frontier}
        \end{equation}
        where \(\mathcal{P}=(d,\Delta)\) denotes a treatment protocol in the
        dose--interval domain and \(\mathrm{TTP}^{\max}_{n_H}(I_0)\) denotes the
        maximum TTP obtained over that domain for the corresponding value of \(n_H\)
        and oxygenation condition. The frontier therefore identifies the set of
        near-optimal treatment protocols for each value of \(n_H\) and oxygenation
        condition. In the following, the explicit dependence on \(I_0\) is omitted for notational simplicity, since all comparisons are performed separately for each oxygenation condition.
        
        The characteristic location of each therapeutic efficiency frontier is summarized through its \emph{TTP-weighted barycenter}, defined as
        \begin{equation}
            \bar d_{n_H}
            =
            \frac{
                \displaystyle
                \sum_{\mathcal{P}\in\mathcal{F}_{n_H}}
                d_{\mathcal{P}}\,
                \mathrm{TTP}_{n_H}(\mathcal{P})
            }{
                \displaystyle
                \sum_{\mathcal{P}\in\mathcal{F}_{n_H}}
                \mathrm{TTP}_{n_H}(\mathcal{P})
            },
            \qquad
            \bar\Delta_{n_H}
            =
            \frac{
                \displaystyle
                \sum_{\mathcal{P}\in\mathcal{F}_{n_H}}
                \Delta_{\mathcal{P}}\,
                \mathrm{TTP}_{n_H}(\mathcal{P})
            }{
                \displaystyle
                \sum_{\mathcal{P}\in\mathcal{F}_{n_H}}
                \mathrm{TTP}_{n_H}(\mathcal{P})
            }.
            \label{eq:S_hill_barycenter}
        \end{equation}
        The barycenter identifies the characteristic location of the near-optimal
        protocols in the dose--interval plane, weighted by their TTP values.
        
        We then compare each alternative Hill coefficient with the reference case
        \(n_H=4\) using three complementary measures. First, we quantify the
        geometrical similarity between the corresponding therapeutic efficiency
        frontiers through the Jaccard index,
        \begin{equation}
            J_{n_H}
            =
            \frac{
                \left|
                \mathcal{F}_{n_H}\cap\mathcal{F}_{4}
                \right|
            }{
                \left|
                \mathcal{F}_{n_H}\cup\mathcal{F}_{4}
                \right|
            }.
            \label{eq:S_hill_jaccard}
        \end{equation}
        Here, the numerator counts the protocols that belong simultaneously to the
        frontier obtained for \(n_H\) and to the reference frontier
        \(\mathcal{F}_4\), whereas the denominator counts all protocols belonging to at
        least one of the two frontiers. Thus, \(J_{n_H}\) measures the fraction of the
        combined near-optimal protocol set that is shared by the two cases. A value of
        \(J_{n_H}=1\) corresponds to identical frontiers, while \(J_{n_H}=0\)
        corresponds to non-overlapping frontiers. In
        Fig.~\ref{Fig.:hill_frontier_sensitivity}, the Jaccard index is reported as a
        percentage for readability; larger percentages therefore indicate greater
        geometrical agreement with the reference case.
        
        Second, changes in the characteristic protocol location are quantified through
        the barycenter shifts
        \begin{equation}
            \delta d_{n_H}
            =
            \bar d_{n_H}-\bar d_4,
            \qquad
            \delta\Delta_{n_H}
            =
            \bar\Delta_{n_H}-\bar\Delta_4.
            \label{eq:S_hill_barycenter_shifts}
        \end{equation}
        These quantities measure whether changes in the Hill coefficient displace the
        high-performing region toward systematically different dose or inter-fraction
        interval regimes.
        
        Finally, we quantify how the predicted TTP changes when the Hill coefficient is
        varied while keeping the treatment protocol fixed. Let \(\mathcal{D}\) denote
        the complete dose--interval protocol domain. As detailed in
        Section~\ref{sec:S_computational_details}, \(\mathcal{D}\) consists of a
        \(50\times50\) grid of dose--interval combinations, corresponding to
        \(2500\) treatment protocols simulated independently. For each
        \(\mathcal{P}\in\mathcal{D}\), we compare the TTP obtained with \(n_H\) with
        the corresponding value obtained for the reference case \(n_H=4\), and
        summarize these differences through
        \begin{equation}
            \mathrm{RMSE}_{n_H}
            =
            \sqrt{
                \frac{1}{|\mathcal{D}|}
                \sum_{\mathcal{P}\in\mathcal{D}}
                \left[
                    \mathrm{TTP}_{n_H}(\mathcal{P})
                    -
                    \mathrm{TTP}_{4}(\mathcal{P})
                \right]^2
            }.
            \label{eq:S_hill_rmse}
        \end{equation}
        The RMSE is expressed in years and measures the typical TTP deviation from the
        reference \(n_H=4\) case over the complete protocol domain.
        
        The results are summarized in
        Fig.~\ref{Fig.:hill_frontier_sensitivity}. A clear separation emerges between \(n_H=1,2\), which produce the largest deviations, and the effectively saturating regime \(n_H\geq3\), for which the results become substantially more stable. This behavior is consistent with the saturation properties of the Hill function discussed above.
                
        The normoxic case is shown in the left panel of
        Fig.~\ref{Fig.:hill_frontier_sensitivity}. The largest changes occur for the
        lowest Hill coefficients. For \(n_H=1\), the frontier overlap with the
        \(n_H=4\) reference is \(78\%\), the barycenter shifts by
        \(-0.083\)~Gy in dose and \(+7.7\)~days in interval, and the TTP deviation
        over the complete protocol domain is \(0.694\)~years. For \(n_H=2\), the
        overlap increases to \(88\%\), while the barycenter shifts are reduced to
        \(-0.050\)~Gy and \(+4.1\)~days and the RMSE decreases to
        \(0.367\)~years. These deviations become substantially smaller once the Hill response enters the saturating regime. For \(n_H=3\), \(5\), and \(6\), the frontier
        overlaps are \(95\%\), \(97\%\), and \(95\%\), respectively, with dose shifts
        between \(-0.012\) and \(+0.006\)~Gy and interval shifts between
        \(-2.0\) and \(+1.6\)~days. The corresponding TTP RMSE values remain small,
        at \(0.146\), \(0.098\), and \(0.168\)~years.
        
        Importantly, the qualitative organization of the normoxic frontier is also
        preserved for \(n_H\geq3\). The near-optimal protocols remain broadly
        distributed in the dose--interval plane, and the reference standard-of-care
        schedule (\(d=1.8\)~Gy, \(\Delta=1\)~day) remains representative of the
        high-performing region. Thus, substantial deviations from the standard-of-care
        schedule do not provide a marked advantage under normoxic conditions. The
        persistence of this pattern across the sufficiently saturating Hill
        coefficients further supports the robustness of the normoxic treatment-selection
        result with respect to the choice of \(n_H\).
        
        The hypoxic case, shown in the right panel of
        Fig.~\ref{Fig.:hill_frontier_sensitivity}, displays the same separation between
        low and sufficiently large Hill coefficients, although the effect of using a
        poorly saturating response is stronger. For \(n_H=1\), the overlap with the
        reference frontier falls to \(15\%\), the barycenter shifts by
        \(+0.910\)~Gy and \(-6.7\)~days, and the TTP RMSE reaches
        \(0.965\)~years. The \(n_H=2\) case is intermediate, with \(57\%\) overlap,
        a dose shift of \(+0.287\)~Gy, an interval shift of only \(-0.1\)~days, and
        an RMSE of \(0.411\)~years. In contrast, the results become substantially more stable for \(n_H\geq3\), although the agreement with the reference case remains slightly weaker than under normoxic conditions: the overlaps are \(88\%\), \(92\%\), and \(90\%\) for \(n_H=3\), \(5\), and \(6\), respectively; the absolute dose shifts do not exceed \(0.036\)~Gy, the interval shifts remain within \(0.8\)~days, and the TTP RMSE values are \(0.115\), \(0.074\), and \(0.105\)~years. 
        
        The hypoxic frontier also preserves the treatment-selection pattern identified
        in the main analysis. For \(n_H\geq3\), its geometry and characteristic
        location remain distinct from the normoxic case and retain the shift toward the
        metronomic protocol regime. Therefore, varying \(n_H\) within the saturating
        range does not remove the qualitative difference between normoxic and hypoxic
        protocol selection. In this sense, the effect associated with the oxygenation
        state remains clearly distinguishable from the comparatively small perturbations
        produced by changing the Hill coefficient within \(n_H\geq3\).
        
        Overall, the sensitivity analysis shows a clear separation between the low-Hill
        cases and the effectively saturating regime. For \(n_H=3,5,6\), the therapeutic
        efficiency frontiers remain close to the \(n_H=4\) reference, their
        TTP-weighted barycenters undergo only small displacements, and the predicted
        TTP values remain quantitatively similar. Most importantly, the characteristic
        difference between the normoxic and hypoxic treatment-selection patterns is
        preserved, including the limited benefit of moving away from the
        standard-of-care schedule under normoxia and the preference for more metronomic
        protocols under hypoxia. The larger deviations observed for \(n_H=1\) and
        \(n_H=2\) are consistent with Hill responses that approach saturation too
        gradually to reproduce the intended oxygen-dependent behavior up to the
        normoxic level \(O_M\). Since \(n_H=4\) was selected to provide an adequately saturating oxygen response over the relevant oxygen range, these results suggest that the main conclusions of the study are not strongly dependent on this specific choice.
        }

    \section*{Supporting figures}

        The following figures provide complementary material supporting the analyses
        presented in the main text and in the preceding sections of the Supporting
        Information.\\

        \begin{figure}[h!]
            \centering
            \captionsetup{font=small}
            \includegraphics[width=\linewidth]{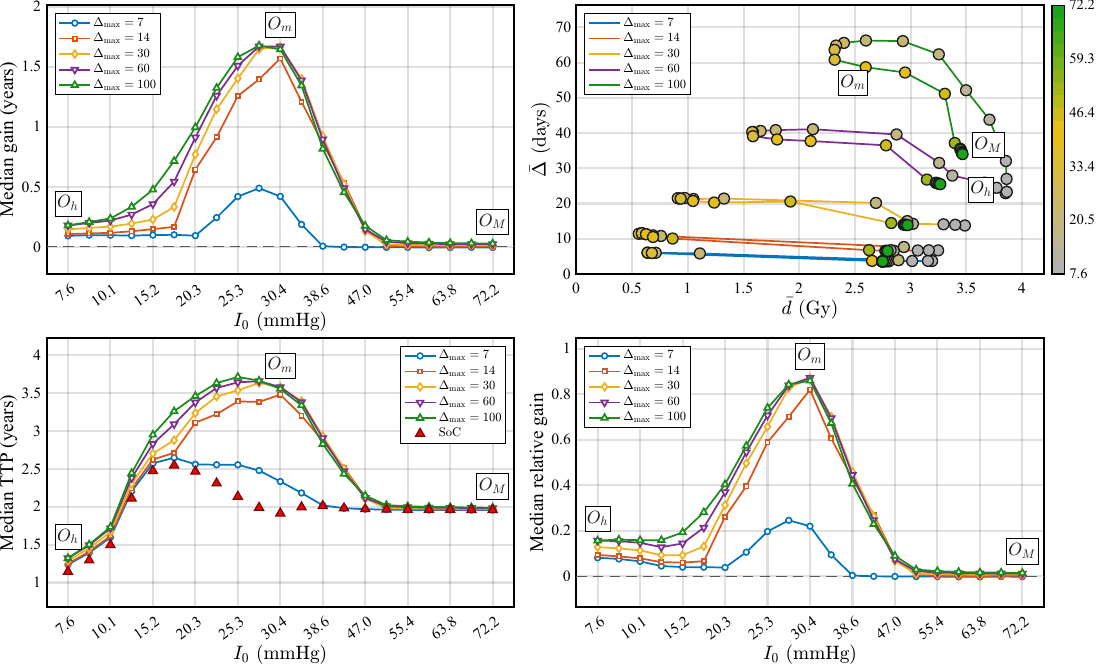}
            \caption{
            Schedule-domain robustness under restricted inter-fraction intervals.
            Dose--interval scans were repeated under the same BED constraint for
            \(\Delta_{\max}\in\{7,14,30,60,100\}\) days, with
            \(\Delta_{\max}=100\) corresponding to the extended range used in the
            main-text dose--interval maps. For each pair \((I_0,\Delta_{\max})\), the
            therapeutic efficiency frontier is recomputed after restricting the admissible
            inter-fraction interval range, and summary statistics are evaluated over the
            corresponding near-optimal protocols.
            \textbf{Top-left:} median TTP gain over SoC.
            \textbf{Top-right:} TTP-weighted frontier barycenters
            \((\bar d,\bar\Delta)\) in the dose--interval plane; lines correspond to
            different values of \(\Delta_{\max}\), while marker colors encode oxygenation
            levels, as indicated by the colorbar.
            \textbf{Bottom-left:} median TTP of near-optimal schedules, together with the
            SoC baseline.
            \textbf{Bottom-right:} median gain normalized by the corresponding SoC TTP
            baseline. Qualitative trends, including the intermediate-hypoxia window and
            barycenter shift, persist as \(\Delta_{\max}\) decreases.
            }
            \label{Fig.:schedule_domain_robustness}
        \end{figure}
    
        \begin{figure}[h!]
            \centering
            \captionsetup{font=small}
            \includegraphics[width=\linewidth]{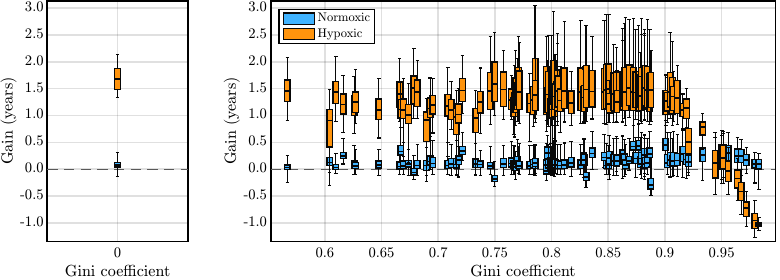}
            \caption{
            Unbinned effect of spatial oxygen-source heterogeneity on protocols selected
            under uniform-supply assumptions. For each \(I_0\in\{O_M,O_m\}\), TTP gains are
            computed as in main-text Fig.~5 for protocols belonging to the uniform-source frontier
            \(\mathcal{F}_u(I_0)\), and evaluated under either the corresponding uniform
            source or heterogeneous oxygen-source configurations, without re-optimization.
            Gains are measured relative to the standard-of-care protocol under the same
            oxygenation regime and spatial configuration. \textbf{Left:} uniform-source reference cases, for which the Gini coefficient is zero. \textbf{Right:} heterogeneous configurations shown individually, without
            pooling them into Gini bins. Each box summarizes the distribution of gains over
            the selected frontier protocols \(\mathcal{P}\in\mathcal{F}_u(I_0)\) for a
            single oxygen-source realization \(V_{O,I_0}^{(m)}(x)\), with spatial
            heterogeneity quantified by the corresponding Gini coefficient
            \(\mathcal{G}_{I_0}^{(m)}\). Blue boxes indicate normoxic conditions
            (\(I_0=O_M\)); orange boxes indicate moderately hypoxic conditions
            (\(I_0=O_m\)). Boxes represent the interquartile range, central lines denote
            medians, and whiskers indicate minimum and maximum gains. The unbinned
            representation confirms the same regime-dependent trend observed in the binned
            analysis, showing that binning is used only to improve readability.}
            \label{Fig.:placeholder}
        \end{figure}

    \begin{figure*}[h!]
        \centering
        \captionsetup{font=small}
        \includegraphics[width=\textwidth]{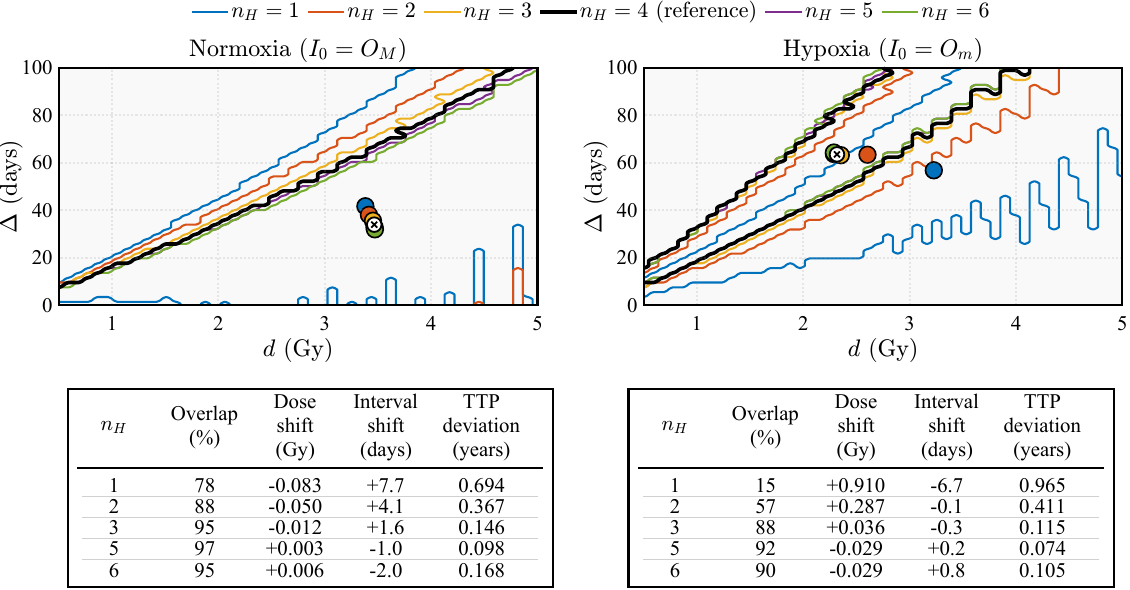}
        \caption{
        Sensitivity of the dose--interval therapeutic efficiency frontier to the Hill
        coefficient \(n_H\).
        Complete dose--interval scans were repeated for
        \(n_H\in\{1,2,3,4,5,6\}\), keeping all other model parameters and simulation
        settings unchanged.
        \textbf{Left:} normoxic condition \(I_0=O_M\).
        \textbf{Right:} hypoxic condition \(I_0=O_m\).
        For each \(n_H\) and oxygenation condition, the therapeutic efficiency
        frontier \(\mathcal{F}_{n_H}\) identifies the near-optimal protocols attaining
        at least \(80\%\) of the corresponding maximum TTP; colored curves show its
        boundary.
        Filled circles indicate the TTP-weighted barycenters
        \((\bar d_{n_H},\bar\Delta_{n_H})\), while the reference case \(n_H=4\) is
        highlighted by the thicker black contour and the white marker with a black
        cross.
        The tables quantify deviations from \(n_H=4\) through the percentage overlap
        between frontiers and the dose and interval shifts of their TTP-weighted
        barycenters, while TTP deviation is quantified by the RMSE (in years),
        computed protocol by protocol over the complete dose--interval domain.
        Overall, for the sufficiently saturating responses \(n_H\geq3\), the
        frontiers retain a high overlap with the reference case, their barycenters
        remain close, and TTP deviations remain limited under both normoxic and
        hypoxic conditions. The larger deviations observed for \(n_H=1\) and
        \(n_H=2\) are associated with Hill profiles that approach saturation more
        gradually over the relevant oxygen range. These results suggest that the
        main treatment-selection conclusions are not strongly dependent on the
        specific reference choice \(n_H=4\).
        }
        \label{Fig.:hill_frontier_sensitivity}
    \end{figure*}
\, 
\newpage
\,
\newpage
\,
\newpage
\,
\newpage
\,

\putbib[SI_references]
\end{bibunit}

\end{document}